\documentclass[preprint,notoc]{JHEP3}

\usepackage{graphicx}
\usepackage{amsmath}
\usepackage{epsf}
\usepackage{psfrag}
\usepackage[sort&compress,numbers]{natbib}

\def\be{\begin{equation}}
\def\ee{\end{equation}}
\def\bea{\begin{eqnarray}}
\def\eea{\end{eqnarray}}

\def\eqn#1{eq.~(\ref{#1})}
\def\eqns#1#2{eqs.~(\ref{#1}) and~(\ref{#2})}

\def\e{\epsilon}

\def\la{\langle}
\def\ra{\rangle}

\def\lr{\leftrightarrow}

\def\Psl{\not{\hbox{\kern-2.3pt $P$}}}
\def\psl{\not{\hbox{\kern-2.3pt $p$}}}
\def\qsl{\not{\hbox{\kern-2.3pt $q$}}}
\def\Ksl{\not{\hbox{\kern-2.3pt $K$}}}
\def\ksl{\not{\hbox{\kern-2.3pt $k$}}}
\def\esl{\not{\hbox{\kern-2.3pt $\pol$}}}

\def\tr{\mathop{\rm tr}\nolimits}
\def\Tr{\mathop{\rm Tr}\nolimits}

\def\pol{\varepsilon}
\def\Neqfour{{\cal N}=4}
\def\Neqone{{\cal N}=1}

\def\spa#1.#2{\left\langle#1\,#2\right\rangle}
\def\spb#1.#2{\left[#1\,#2\right]}
\def\lor#1.#2{\left(#1\,#2\right)}
\def\sand#1.#2.#3{%
\left\langle\smash{#1}{\vphantom1}^{-}\right|{#2}%
\left|\smash{#3}{\vphantom1}^{-}\right\rangle}
\def\sandp#1.#2.#3{%
\left\langle\smash{#1}{\vphantom1}^{-}\right|{#2}%
\left|\smash{#3}{\vphantom1}^{+}\right\rangle}
\def\sandpp#1.#2.#3{%
\left\langle\smash{#1}{\vphantom1}^{+}\right|{#2}%
\left|\smash{#3}{\vphantom1}^{+}\right\rangle}
\def\sandpm#1.#2.#3{%
\left\langle\smash{#1}{\vphantom1}^{+}\right|{#2}%
\left|\smash{#3}{\vphantom1}^{-}\right\rangle}
\def\sandmp#1.#2.#3{%
\left\langle\smash{#1}{\vphantom1}^{-}\right|{#2}%
\left|\smash{#3}{\vphantom1}^{+}\right\rangle}
\def\sandmm#1.#2.#3{%
\left\langle\smash{#1}{\vphantom1}^{-}\right|{\slash\!\!\! #2}%
\left|\smash{#3}{\vphantom1}^{-}\right\rangle}
\def\spab#1.#2.#3{\sandmm#1.#2.#3}
\def\spbb#1.#2.#3.#4{\sandpm#1.{\slash\!\!\! #2\slash\!\!\! #3}.#4}
\newbox\charbox
\newbox\slabox
\def\s#1{{      
        \setbox\charbox=\hbox{$#1$}
        \setbox\slabox=\hbox{$/$}
        \dimen\charbox=\ht\slabox
        \advance\dimen\charbox by -\dp\slabox
        \advance\dimen\charbox by -\ht\charbox
        \advance\dimen\charbox by \dp\charbox
        \divide\dimen\charbox by 2
        \raise-\dimen\charbox\hbox to \wd\charbox{\hss/\hss}
        \llap{$#1$}
}}
\def\ksl{\s{k}}

\newcommand{\EQ}[1]{\begin{equation} #1 \end{equation}}

\newcommand{\SP}[1]{\begin{equation}\begin{split} #1 \end{split}\end{equation}}

\def\beqa{\begin{eqnarray}}
\def\eeqa{\end{eqnarray}}
\def\beq{\begin{equation}}
\def\eeq{\end{equation}}
\def\hf{{\textstyle{1\over2}}}
\def\ihf{{\textstyle{i\over2}}}
\def\sst{\scriptscriptstyle}

\def\vev#1{\langle{#1}\rangle}

\def\MHVb{$\overline{\text{MHV}}$}

\preprint{
  hep-th/0412275\\
  IPPP/04/88\\
  DCPT/04/176\\
  December, 2004}

\title{MHV Rules for Higgs Plus Multi-Parton Amplitudes}

\author{S.D. Badger, 
    \ E. W. N. Glover and Valentin V. Khoze\\
        Institute of Particle Physics Phenomenology,
        Department of Physics,\\
        University of Durham, Durham, DH1 3LE, UK\\
        E-mail: \email{s.d.badger@durham.ac.uk, 
	e.w.n.glover@durham.ac.uk, 
        valya.khoze@durham.ac.uk}.
}

\abstract{
We present MHV-rules for 
constructing perturbative amplitudes for a Higgs boson and an arbitrary number of 
partons.  
We give explicit expressions for amplitudes involving a Higgs and three negative helicity
partons and any number of positive helicity partons - the NMHV amplitudes.
We also present a recursive formulation of MHV rules that incorporates the Higgs, quarks and
gluons.   The recursion relations are valid for all non-MHV amplitudes.
The general results agree numerically with all of the available
Higgs $+$ $n$-parton
amplitudes and in some cases provide considerably shorter expressions. }


\dedicated{\centerline{Submitted to JHEP}}

\begin{document}
\section{Introduction}

There is a new and poweful framework for computing gauge theory amplitudes --
`MHV rules'~\cite{CSW1}. These rules provide analternative and more compact diagrammatic
expansion compared to the conventional Feynman rules.
The basic building blocks are the colour-ordered $n$-point vertices which are connected by
scalar propagators. These vertices are off-shell continuations of the maximally
helicity-violating (MHV) $n$-gluon scattering amplitudes  of Parke and
Taylor~\cite{ParkeTaylor}. They contain precisely two negative helicity gluons.
By connecting MHV vertices,  amplitudes involving
more negative helicity gluons can be built up. The MHV rules are inspired by
the interpretation of supersymmetric Yang-Mills (SYM) theory and
QCD as a topological string propagating in twistor space~\cite{Witten1}.  

The original MHV rules for gluons~\cite{CSW1} have been extended
to amplitudes with fermions~\cite{GK}. In fact, they straighforwardly apply to
all particles in   the full $\Neqfour$ supermultiplet by reading the appropriate
terms from the general expression for the MHV ${\cal N}=4$ supervertex of Nair
\cite{Nair} using the algorithm described in \cite{GGK,VVK}.

MHV rules have been shown to work at the loop-level for supersymmetric
theories and, building on the earlier work of Bern et al~\cite{BDDK1,BDDK2},
 there has been enormous progress in computing multi-leg
loop amplitudes in $\Neqfour$~\cite{CSW2,BST,CSW3,BBKR,Cachazo,
BCF1,BDDK7,BCF2,BCF3,BDKNMHV} 
and $\Neqone$~\cite{QuigleyRozali,BBST1,BBDD,BBDW}
SYM  theories.\footnote{Progress has also been made for non-supersymmetric loop amplitudes~\cite{BBST2}}

At the same time, the MHV rules have given new tree-level gauge-theory results for non-MHV
amplitudes  involving gluons~\cite{Zhu,KosowerNMHV,BBK}, and fermions~\cite{GK,GGK,Wu1,Wu2}. It is of
great phenomenological interest to extend the method to processes involving massive
particle like the Higgs~\cite{DGK}, electroweak vector bosons~\cite{BFKM} and heavy
quarks.  Processes like these, together with  those involving massive supersymmetric
particles, are of vital importance for experiments at the LHC and elsewhere.  

It has been shown in \cite{DGK} that  Higgs plus gluons amplitudes can be computed
efficiently by introducing an effective operator and using a modified set of MHV rules.
Approximating Higgs interactions by coupling via a top loop is strongly  supported by
current precision electroweak data which predicts the the Higgs is considerably lighter
than  $2m_t \approx 360$~GeV; currently $m_H < 260$~GeV at 95\% confidence
level~\cite{Renton:2004wd}. In this case it is an excellent approximation
to integrate out the heavy top quark, summarising its effects via the dimension-five
operator  $H \tr G_{\mu\nu}
G^{\mu\nu}$~\cite{HggOperator2,HggOperator5}. This
operator can then be `dressed' by standard QCD vertices in order  to generate Higgs plus
multi-parton amplitudes.

The purpose of this paper is to complete the work begun in \cite{DGK} by computing Higgs to
multi-parton amplitudes with one and two quark-antiquark pairs. In section \ref{sec:tower}
we review the construction of the amplitudes for Higgs and gluons. The MHV rules arise from
splitting the effective Higgs-gluons interaction into two pieces, one holomorphic, and one
anti-holomorphic. These two pieces can be calculated separately using scalar MHV graphs and
then must be added together to form the full Higgs amplitude.  Here we generalise this
method to allow fermions in the Higgs plus multi-parton amplitudes and show how MHV and
\MHVb\ vertices for  one and two quark-antiquark pairs can be derived from supersymmetry
and  extend the MHV rules for Higgs and gluons to the case including fermions.  

In Section \ref{sec:2q} we  write down the MHV amplitudes for a self-dual scalar $\phi$
into a quark-antiquark pair and gluons and then use these to compute the corresponding NMHV
amplitudes. We then show that these results agree with the known analytic expressions for
$n\le5$ partons at tree-level.  In Section \ref{sec:4q} we consider interactions containing
two quark-anti-quark pairs. We use the new MHV vertices to compute the eight NMHV helicity
amplitudes which are then shown to reproduce the known analytic formula for $n\le5$
partons. 

Finally Section \ref{sec:recurs} generalises the recursive formulation of non-MHV gluonic
amplitudes of Bena, Bern and Kosower~\cite{BBK} to the case of amplitudes with fermions
and/or scalars. We then use these formulae to find numerical values for  NNMHV amplitudes
with Higgs and a single quark-antiquark pair. This completes the checks necessary for  all
amplitudes with up to $n\le5$ partons. The recursion relations are, of course, valid for
$n> 5$ partons. Our findings are then summarised in the conclusions.  

For the sake of clarity and completeness, we include three appendices. Appendix A defines
our conventions regarding helicity spinor formalism. Appendix B defines colour ordered
amplitudes which we study in this paper. Finally Appendix C provides a recursive proof of
the vanishing of all $A_n(\phi, q^\lambda,g^+_2,\ldots,g^+_{n-1},\bar{q}^{-\lambda}_n)$
amplitudes using the Berends-Giele currents \cite{BerendsGiele}.

\section{MHV amplitude towers including fermions \label{sec:tower}}

 The MHV rules arise
from splitting the effective Higgs-gluons interaction into two pieces, one
holomorphic, and one anti-holomorphic. The amplitudes generated by these two terms
should be calculated
separately using scalar MHV graphs and then must be added together to form the full Higgs amplitude.
In this way the picture \cite{DGK} of the ``two towers" of MHV and \MHVb~amplitudes to form the full
Higgs amplitude emerges.

\subsection{The model}
\label{ModelSection}

In the Standard Model the Higgs boson couples to gluons through
a fermion loop. The dominant contribution is from the top quark.
For large $m_t$, the top quark can be integrated out leading
to the effective interaction~\cite{HggOperator5,HggOperator2},
 \be
 {\cal L}_{\sst H}^{\rm int} =
  {C\over2}\, H \tr G_{\mu\nu}\, G^{\mu\nu}  \ .
 \label{HGGeff}
 \ee
In the Standard Model, and to leading order in $\alpha_s$,
the strength of the interaction is given by $C = \alpha_s/(6\pi v)$,
with $v = 246$~GeV.

It was explained in \cite{DGK} that
the MHV or twistor-space structure of the Higgs-plus-gluons amplitudes is best
elucidated by dividing the Higgs coupling to gluons, \eqn{HGGeff}, into
two terms, containing purely selfdual (SD) and purely anti-selfdual (ASD)
gluon field strengths,
\be
G_{\sst SD}^{\mu\nu} = \hf(G^{\mu\nu}+{}^*G^{\mu\nu}) \ , \quad
G_{\sst ASD}^{\mu\nu} = \hf(G^{\mu\nu}-{}^*G^{\mu\nu}) \ , \quad
{}^*G^{\mu\nu} \equiv \ihf \epsilon^{\mu\nu\rho\sigma} G_{\rho\sigma} \ .
\ee
This division can be accomplished by considering $H$ to be the real
part of a complex field $\phi = {1\over2}( H + i A )$, so that
\bea
 {\cal L}^{\rm int}_{H,A} &=&
{C\over2} \Bigl[ H \tr G_{\mu\nu}\, G^{\mu\nu}
             + i A \tr G_{\mu\nu}\, {}^*G^{\mu\nu} \Bigr]
 \label{effinta}\\
&=&
C \Bigl[ \phi \tr G_{{\sst SD}\,\mu\nu}\, G_{\sst SD}^{\mu\nu}
 + \phi^\dagger \tr G_{{\sst ASD}\,\mu\nu} \,G_{\sst ASD}^{\mu\nu} \Bigr]
 \ .
 \label{effintb}
\eea
The important observation of \cite{DGK} was that, due to selfduality, the amplitudes for
$\phi$ plus $n$ gluons, and those for $\phi^\dagger$ plus $n$ gluons,
separately have a simpler structure than the amplitudes for
$H$ plus $n$ gluons. MHV rules are constructed for $\phi$ plus $n$ gluons amplitudes
and for for $\phi^\dagger$ plus $n$ gluons separately.
Then since $H = \phi + \phi^\dagger$,
the Higgs amplitudes can be recovered
as the sum of the $\phi$ and $\phi^\dagger$ amplitudes.

Quarks do not enter the $HGG$ effective vertex \eqref{HGGeff} or \eqref{effintb} directly,
they couple to it only through gluons. The division of \eqn{HGGeff} into selfdual
and anti-selfdual terms, dictated by \eqn{effintb}  will continue to
be the guiding principle for constructing MHV rules for the Higgs plus quarks and gluons
amplitudes. In fact, in this section we will {\it derive}
the MHV rules for the Higgs with gluons and quarks
from the simpler MHV rules
of \cite{DGK} for amplitudes with the Higgs and gluons only.

Throughout the paper
we use a standard colour decomposition for all amplitudes ${\cal A}_n $
in terms of colour factors ${\cal T}_n$
and purely kinematic partial
amplitudes $A_n$ as reviewed in Appendix~\ref{Cdecomp}.
Only these colour-ordered kinematic amplitudes $A_n$ need to be calculated.
The full amplitudes
can be assembled from color-ordered amplitudes $A_n$
and known expressions for  the colour factors ${\cal T}_n$ as
described in Appendix~\ref{Cdecomp}.

Hence without loss of generality from now on we will concentrate on the colour-ordered
partial amplitudes $A_n$ involving partons (gluons, quark-antiquark pairs, gluinos)
plus the single colourless scalar field which can be
$H$, $\phi$ or $\phi^\dagger$.

The kinematic amplitudes $A_{n}$ have the colour information stripped off
and hence do not distinguish between fundamental quarks and adjoint gluinos.
Hence, if we know kinematic amplitudes involving gluinos in a supersymmetric theory,
we automatically know kinematic amplitudes with quarks,
\be
A_{n}(H,q^{+},\ldots,\bar{q}^{\,-},\ldots, g^+,\ldots, g^-)\, = \,
A_{n}(H,\Lambda^+,\ldots,\Lambda^-,\ldots, g^+,\ldots, g^-)
\label{four-1}
\ee
Here $q^{\pm}$, $\bar{q}^{\pm}$, $g^{\pm}$, $\Lambda^{\pm}$ denote
quarks, antiquarks, gluons and gluinos of $\pm$ helicity, and
$H$ represents the colourless scalar $H$, $\phi$, $\phi^\dagger$ or simply nothing.
By this we mean that
\eqref{four-1} is valid with or without the scalar field, this is because
the colourless scalar does not modify the colour decomposition.
We conclude from \eqref{four-1} that knowing kinematic amplitudes in a supersymmetric
theory with gluinos allows us to deduce immediately non-supersymmetric amplitudes with quarks
and antiquarks.

MHV rules were formulated in \cite{DGK} for amplitudes with the Higgs and $n$
gluons. In the following section we will show that these rules uniquely
determine the MHV rules for the Higgs plus all partons (i.e. gluons and
quarks). More precisely, the MHV amplitudes with the Higgs and gluons determine
the MHV amplitudes with the Higgs, gluinos and gluons via supersymmetric Ward
identities. Then \eqref{four-1} turns gluinos into quarks in a
non-supersymmetric theory.

\subsection{MHV amplitudes}

We start with the (anti)-MHV amplitudes for the Higgs and gluons from \cite{DGK}.
First, the decomposition of the $HGG$ vertex into the selfdual and the anti-selfdual
terms \eqn{effintb}, guarantees that the whole class of helicity amplitudes must vanish
\cite{DGK}:
\be
  A_n(\phi,g_1^\pm,g_2^+,g_3^+,\ldots,g_n^+) = 0 \, , \quad
  A_n(\phi^\dagger,g_1^\pm,g_2^-,g_3^-,\ldots,g_n^-) = 0 \, ,
 \label{phimpvanish}
\ee
for all $n$.

The amplitudes, with precisely two negative helicities,
$\phi\, g^- g^+ \ldots g^+ g^- g^+ \ldots  g^+$, are the
first non-vanishing $\phi$ amplitudes. These amplitudes will be referred to as
the $\phi$-MHV amplitudes.
General factorization properties now imply that they have to be extremely
simple \cite{DGK}, they read
\be
A_n(\phi,g_1^+,g_2^+,\ldots, g_p^-, \ldots, g_q^-, \ldots ,g_n^+) =
 { {\spa{p}.{q}}^4 \over \spa1.2 \spa2.3 \cdots \spa{n-1,}.{n} \spa{n}.1 } \,,
\label{assrtn}
\ee
Here only legs $p$ and $q$ have negative helicity.
This expression is valid for all $n$.
Besides the correct collinear and multi-particle factorization
behavior, these amplitudes also correctly reduce to pure QCD
MHV amplitudes as the $\phi$ momentum approaches zero. In fact, the expressions
\eqref{assrtn} for $\phi$-MHV $n$-gluon amplitudes are exactly the same
as the MHV $n$-gluon amplitudes in pure QCD. The only difference of
\eqref{assrtn} with pure QCD is that the total momentum carried by gluons,
$p_1+p_2+\ldots+p_n=-p_{\phi}$ is the momentum
carried by the $\phi$-field and is non-zero.
This momentum makes the Higgs case well-defined
on-shell for fewer legs than in the pure QCD case.
The first few $\phi$ amplitudes have the form,
\bea
  A_2(\phi,g_1^-,g_2^-) &=& { {\spa1.2}^4 \over \spa1.2 \spa2.1 } = - {\spa1.2}^2,
 \label{Hmm} \\
  A_3(\phi,g_1^-,g_2^-,g_3^+) &=& { {\spa1.2}^4 \over \spa1.2 \spa2.3 \spa3.1 }
                    = { {\spa1.2}^3 \over \spa2.3 \spa3.1 } \,,
 \label{Hmmp} \\
 A_4(\phi,g_1^-,g_2^-,g_3^+,g_4^+) &=& { {\spa1.2}^4 \over \spa1.2\spa2.3\spa3.4\spa4.1 }
                   \,.
 \label{Hmmpp}
\eea

Since the MHV amplitudes~(\ref{assrtn}) have an identical form to the
corresponding amplitudes of pure Yang-Mills theory,
\cite{DGK} argued that their off-shell continuation should also identical to
that proposed in the pure-glue context in~\cite{CSW1}.
Everywhere the off-shell leg $i$ carrying momentum $P_i$ appears
in \eqref{assrtn}, we let the corresponding holomorphic spinor be
$\lambda_{i,\alpha} = (P_i)_{\alpha\dot\alpha} \xi^{\dot\alpha}$.
Here $\xi^{\dot\alpha}$ is an arbitrary reference spinor, chosen to be
the same for all MHV diagrams contributing to the amplitude.

We are now ready to discuss MHV amplitudes with gluons and fermions. To this end we
first consider a pure ${\cal N}=1$ supersymmetric Yang-Mills (without the Higgs or quarks).
An MHV amplitude $A_n=A_{l+2m}$
with $l$ gluons, $g$, and $2m$ gluinos, $\Lambda$, in the ${\cal N} = 1$ pure gauge theory
exists only for $m=0,1,2$.
This is because it must have precisely $n-2$ particles with positive helicity
and
$2$ with negative helicity, and gluinos always come in pairs
with helicities $\pm {1\over 2}$.
Hence, there are
three types of MHV tree amplitudes in the ${\cal N} = 1$ pure gauge theory:
\be
\label{threecls}
A_n (g_p^-,g_q^-) \ , \quad
A_n (g_t^-,\Lambda_r^-,\Lambda_s^+)\ , \quad
A_n (\Lambda_t^-,\Lambda_s^+,\Lambda_r^-,\Lambda_q^+) \ .
\ee
The MHV purely gluonic amplitude is \cite{ParkeTaylor,BerendsGiele}:
\be
A_n (g_p^-,g_q^-)=\
{\vev{p~q}^4 \over \prod_{i=1}^n \vev{i~i+1}} \ ,
\label{mpng}
\ee
where $\lambda_{n+1} \equiv \lambda_1$. For notational simplicity in this and the following
expressions for MHV amplitudes we do not show explicitly the positive helicity gluons
$g^+$.
The MHV amplitude with two external fermions and $n-2$ gluons is
\be
\label{ndcls}
A_n (g_t^-,\Lambda_r^-,\Lambda_s^+)= \
{\vev{t~r}^3\ \vev{t~s} \over \prod_{i=1}^n \vev{i~i+1}} \ , \quad
A_n (g_t^-,\Lambda_s^+,\Lambda_r^-)= \
-\ {\vev{t~r}^3\ \vev{t~s} \over \prod_{i=1}^n \vev{i~i+1}} \ ,
\ee
where the first expression corresponds to $r<s$ and the second to $s<r$
(and $t$ is arbitrary).
The MHV amplitudes with four fermions and $n-4$ gluons on external lines are
\be
\label{ndcls2}
A_n (\Lambda_t^-,\Lambda_s^+,\Lambda_r^-,\Lambda_q^+)
= \
{\vev{t~r}^3\ \vev{s~q} \over \prod_{i=1}^n \vev{i~i+1}} \ , \quad
A_n (\Lambda_t^-,\Lambda_r^-,\Lambda_s^+,\Lambda_q^+)
= \
-\ {\vev{t~r}^3\ \vev{s~q} \over \prod_{i=1}^n \vev{i~i+1}}
 \
\ee
The first expression in \eqref{ndcls2}
corresponds to $t<s<r<q,$ the second -- to $t<r<s<q,$
and there are other similar expressions,
obtained by further permutations of fermions, with the overall
sign determined by the ordering.

We now recall that expressions \eqref{ndcls}, \eqref{ndcls2} are not independent inputs
into the MHV programme, they follow from the amplitudes \eqref{mpng} via
supersymmetric
Ward identities \cite{SWI1,SWI2,SWI3,SpinorHelicity6,LDTASI}.

Supersymmetric Ward identities
\cite{SWI1,SWI2,SWI3} follow from the fact that,
supercharges $Q$ annihilate the vacuum,
and hence we have an equation,
\EQ{
\langle [Q\, , \,
\Lambda^+_k \ldots g_{r_1}^- \ldots g_{r_2}^- \ldots]\rangle
\ = \ 0
\ , }
where dots indicate positive helicity gluons. In order to make
anticommuting spinor $Q$ to be a singlet entering a commutative
(rather than anticommutative) algebra
with all the fields we contract it with a commuting spinor $\eta$ and
multiply it by a Grassmann number $\theta$. This defines a
commuting singlet operator $Q(\eta).$
Following \cite{LDTASI} we can write down the following susy algebra relations,
\SP{\label{susyward}
[Q(\eta) \, , \, \Lambda^{+}(k)] \ = \ - \theta \vev{\eta~k}\,g^+ (k) \ , \quad
[Q(\eta) \, , \, \Lambda^{-}(k)] \ = \ + \theta [\eta~k]\,g^- (k) \ , \\
[Q(\eta) \, , \, g^{-}(k)] \ = \ + \theta \vev{\eta~k}\,\Lambda^- (k) \ , \quad
[Q(\eta) \, , \, g^{+}(k)] \ = \ - \theta [\eta~k]\,\Lambda^+ (k) \ .
}
In what follows, the anticommuting parameter
$\theta$ will cancel from the relevant expressions for the amplitudes.
The arbitrary spinors  $\eta_a,$ $\eta_{\dot a},$  will be fixed below.
It then follows from \eqref{susyward} that
\be
\label{restggg}
\theta {\vev{\eta~k}}\, A_n (g_{r_1}^- , g_{r_2}^- ) = \
-\theta {\vev{\eta~r_1}}\,
A_n (\Lambda_{k}^+, \Lambda_{r_1}^-, g_{r_2}^-)
- \,\theta
{\vev{\eta~r_2}}\,
A_n (\Lambda_{k}^+, g_{r_1}^-, \Lambda_{r_2}^-)
 \ .
\ee
The minus signs on the right hand side arise from anticommuting $\theta$ with
gluino fields.
After cancelling $\theta$ and choosing $\eta$ to be one of the two $r_j$ we find from
\eqref{restggg} that the purely gluonic amplitude is proportional to the
amplitude with two gluinos,
\be
\label{restggg2}
 A_n (g_{r_1}^- , g_{r_2}^- ) \ = \
-\,{{\vev{r_2~r_1}} \over {\vev{r_2~k}}} \,
A_n (\Lambda_{k}^+, \Lambda_{r_1}^-, g_{r_2}^-) \ = \
-\,{{\vev{r_1~r_2}} \over {\vev{r_1~k}}}
\,
A_n (\Lambda_{k}^+, g_{r_1}^-, \Lambda_{r_2}^-)
 \ .
\ee
This gives the MHV amplitudes \eqref{ndcls}.
Equations \eqref{ndcls2} follow from a similar construction.

We can now add the Higgs scalars $\phi$ and $\phi^\dagger$ to the construction of MHV
amplitudes above. To achieve this, we embed the bosonic effective interaction
\eqref{effintb} into a pseudo-supersymmetric theory,
\be
{\cal L}^{\rm int} =
- C \int d^2\theta\  \phi \,\tr W^\alpha W_\alpha \,
- C \, \int d^2\bar\theta\ \phi^{\dagger}\,
 \tr \overline{W}_{\dot\alpha}\overline{W}^{\dot\alpha} \,.
\label{susyembd}
\ee
Here $G_{\sst SD}^{\mu\nu}$ is the bosonic component of the ${\cal N}=1$ chiral
superfield $W_\alpha(x,\theta)$, but $\phi$ is not a superfield, it is still
a (single component) scalar field $\phi(x)$ which has no superpartners.
So, the theory described by \eqref{susyembd} is not a supersymmetric theory.
However, there is a continuous symmetry group which leaves this action
invariant. It is generated by the `supercharges' $Q$ which act non-trivially
on gluons and gluinos -- precisely as in \eqref{susyward} -- and at the same time
annihilate the scalar field,
\be \label{susywardphi}
[Q(\eta) \, , \, \phi(p)] \ = \ 0 \ , \qquad
[Q(\eta) \, , \, \phi^\dagger (p)] \ = \ 0 \ .
\ee

Applying the commutation relations, \eqref{susyembd}, \eqref{susywardphi} to
equation
\EQ{
\langle [Q\, , \,
\phi \,\Lambda^+_k \ldots g_{r_1}^- \ldots g_{r_2}^- \ldots]\rangle
\ = \ 0
\ , }
we find the same relation as in \eqn{restggg2}, but now for the MHV amplitudes with
the Higgs field $\phi$.

We conclude that the from the fact that the purely gluonic MHV amplitudes,
\eqref{assrtn} and \eqref{mpng} take the same form,
the pseudo-supersymmetry Ward identities guarantee that the tree-level
$\phi$-MHV amplitudes with fermions and gluons have exactly the same algebraic form
as the corresponding MHV amplitudes in pure ${\cal N}=1$ gauge theory,
\eqref{ndcls} and \eqref{ndcls2}.
Hence we now can insert the $\phi$ field on the left hand sides of
\eqref{ndcls} and \eqref{ndcls2}, and, furthermore, replace gluinos
with quarks as in \eqref{four-1}.

We need to be slightly more careful in order to deduce the $\phi$-MHV amplitudes with
two quark-antiquark pairs of different flavours, i.e.
${A}_n(\phi,q^+,\bar{q}^{\,-},Q^+,\bar{Q}^-)$ where $q$ and $Q$ denote the two different
quarks. Such amplitudes are obtained from the ${\cal N}=2$ supersymmetric amplitudes
${A}_n(\phi,\Lambda_{(1)}^+,\Lambda_{(1)}^-,\Lambda_{(2)}^+,\Lambda_{(2)}^-)$ where
$\Lambda_{(1)}$ and $\Lambda_{(2)}$ are gluinos from two different ${\cal N}=1$
supermultiplets.
All such amplitudes can be read off from the general expression for the MHV ${\cal N}=4$
supervertex of Nair \cite{Nair} using the algorithm described in \cite{GGK,VVK}. The
supervertex and the corresponding component vertices comply with the supersymmetric
Ward identities in pure ${\cal N}=4$ theory. The Higgs field $\phi$ can always be
added to these amplitudes in precisely the same way as above, without changing the
expression for the vertex. This follows from promoting the ${\cal N}=4$ or
${\cal N}=2$ supersymmetry to a `pseudo' supersymmetry by augmenting the algebra
with the condition \eqref{susywardphi}.

\subsection{MHV rules}

We have argued that
the complete set of MHV amplitudes in QCD coupled to the Higgs field
consists of $n$-parton amplitudes made out of one or less scalar field $\phi$,
an arbitrary number of gluons and $m=0,1,2$ quark-antiquark pairs. All these amplitudes
have precisely two negative helicities. Schematically, they are
\bea
\label{phimhvamps}
A_n(\phi,g_p^-,g_q^-) \ , \quad
A_n(\phi,q_1^{-\lambda},g_r^-,\bar{q}_n^{\,\lambda}) \ , \quad
A_n(\phi,q_1^{-\lambda_1},\bar{Q}_s^{\,\lambda_2},Q_{s+1}^{-\lambda_2},\bar{q}_n^{\,\lambda_1})\ , \\
\label{mhvamps}
A_n(g_p^-,g_q^-) \ , \quad
A_n(q_1^{-\lambda},g_r^-,\bar{q}_n^{\,\lambda}) \ , \quad
A_n(q_1^{-\lambda_1},\bar{Q}_s^{\,\lambda_2},Q_{s+1}^{-\lambda_2},\bar{q}_n^{\,\lambda_1}) \ .
\eea
Here we have not shown the positive helicity gluons and did not exhibit all different orderings
for amplitudes with two quark-antiquark pairs. As before, $Q^{-\lambda_2}$ and $\bar{Q}^{\,\lambda_2}$
denote the second flavour of (anti)quarks with helicities $\pm \lambda_2$.

The first line, \eqn{phimhvamps}, gives the $\phi$-MHV amplitudes, and the second line,
\eqn{mhvamps} corresponds to standard QCD MHV amplitudes.
The \MHVb~amplitudes are obtained from (\eqn{phimhvamps}) and (\eqn{mhvamps}) by parity.
They are
\bea
\label{phimhvbamps}
A_n(\phi^\dagger,g_p^+,g_q^+) \ , \quad
A_n(\phi^\dagger,q_1^{\lambda},g_r^+,\bar{q}_n^{\,-\lambda}) \ , \quad
A_n(\phi^\dagger,q_1^{\lambda_1},\bar{Q}_s^{\,-\lambda_2},Q_{s+1}^{\lambda_2},\bar{q}_n^{\,-\lambda_1})\ , \\
\label{mhvbamps}
A_n(g_p^+,g_q^+) \ , \quad
A_n(q_1^{\lambda},g_r^+,\bar{q}_n^{\,-\lambda}) \ , \quad
A_n(q_1^{\lambda_1},\bar{Q}_s^{\,-\lambda_2},Q_{s+1}^{\lambda_2},\bar{q}_n^{\,-\lambda_1}) \ ,
\eea
where we have not shown the negative helicity gluons.

We start with the amplitudes with no fermions, considered in \cite{DGK}.
The left (red) tower in figure 1 lays out the MHV structure of the $\phi$ plus
multi-gluon amplitudes.   All non-vanishing amplitudes are labelled with
circles.  The  fundamental $\phi$-MHV vertices, which coincide with the $\phi
g^-g^- g^+ \ldots g^+$ amplitudes, are the basic building
blocks and are labelled by red dots. The result of combining
$\phi$-MHV vertices with pure-gauge-theory MHV vertices is to produce
amplitudes with more than two negative helicities. These amplitudes
are represented as red open circles.  Each MHV diagram contains
exactly one $\phi$-MHV vertex; the rest are pure-gauge-theory MHV
vertices.  The vertices are combined with scalar propagators.
The MHV-drift is always to the left and upwards.  Collectively, these
amplitudes form the holomorphic (or MHV) tower of accessible amplitudes.

\begin{figure}[t]
    \begin{center}
        \includegraphics[width=8cm]{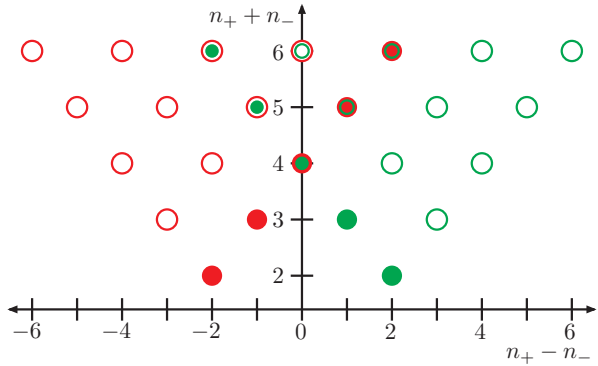}
    \end{center}
    \caption{The structure of Higgs plus multi-gluon amplitudes obtained by
combining the MHV tower for $\phi+n$~gluons and the anti-MHV tower
of $\phi^\dagger +n$~gluon amplitudes.}
    \label{TreeMapFigure}
\end{figure}

The corresponding amplitudes for $\phi^\dagger$ are shown in
the right (green) tower in figure \ref{TreeMapFigure}. They can be obtained by applying parity to the
$\phi$ amplitudes. For practical purposes this means that we compute with
$\phi$, and reverse the helicities of every gluon. Then we let
$\spa{i}.{j} \lr \spb{j}.{i}$ to get the desired $\phi^\dagger$
amplitude. The set of building-block amplitudes are therefore anti-MHV.
Furthermore, the amplitudes with additional positive-helicity gluons are
obtained by combining with anti-MHV gauge theory vertices. The
anti-MHV-drift is always to the right and upwards.  Collectively, these
amplitudes form the anti-holomorphic (or anti-MHV) tower of accessible
amplitudes.

The allowed helicity states are shown in figure \ref{TreeMapFigure}
and are composed of both holomorphic and anti-holomorphic structures.
Where the two towers do not overlap, the 
amplitudes for the real Higgs boson with gluons coincide with the
$\phi$ ($\phi^\dagger$) amplitudes.   On the other hand, where the towers overlap,
we add the $\phi$ and $\phi^\dagger$ amplitudes.

Next we consider amplitudes with one quark-antiquark pair.
Helicity must be conserved along the quark line so the all
plus configuration is trivially zero. The case where antiquark has opposite helicity to the quark
and all gluons have positive helicity is also zero, see Appendix \ref{app:vanish} for a proof
which does not appeal to supersymmetry.
So the first non-vanishing amplitude is again with two negative helicities and
one of them lying on a
gluon: $A_n(\phi,q^{\lambda}_1,\ldots,g^-_s,\ldots,\bar{q}^{\,-\lambda}_n)$. This is
precisely the second $\phi$-MHV amplitude in \eqn{phimhvbamps}.

\begin{figure}[t]
    \begin{center}
        \includegraphics[width=8cm]{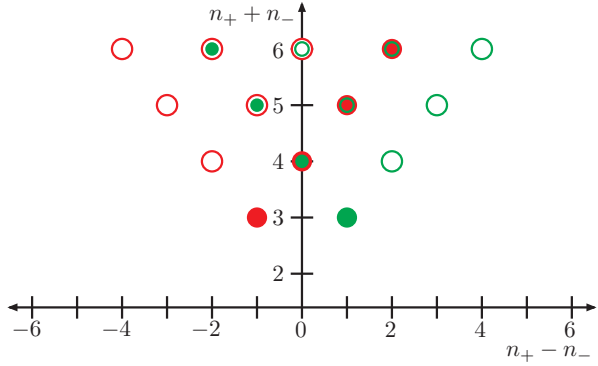}
    \end{center}
    \caption{
    The structure of Higgs plus multi-gluon plus quark-antiquark pair amplitudes obtained by
combining the MHV tower for $\phi+q\bar q+n$~gluons  and the anti-MHV tower
of $\phi^\dagger +q\bar q+n$~gluon amplitudes.}
    \label{fig:treemap}
\end{figure}

The structure of the two MHV towers for amplitudes with the Higgs and one quark-antiquark pair
is set out in figure \ref{fig:treemap}. Here the $\phi$-MHV
amplitudes are represented by filled red dots and the
$\phi$-\MHVb~by filled green dots. The open red(green) dots are amplitudes which can be found
by combining two or more MHV(\MHVb) vertices. The \MHVb~amplitudes can be obtained directly
from the MHV amplitudes via parity transformation. Once again,
Higgs amplitudes are given directly by $\phi$ or $\phi^\dagger$ amplitudes, 
or by adding them when the towers overlap.

\begin{figure}[t]
    \begin{center}
        \includegraphics[width=8cm]{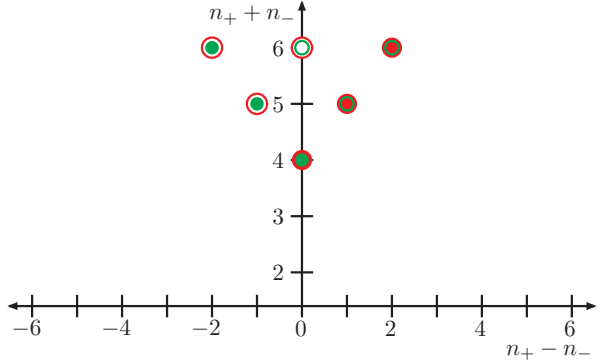}
    \end{center}
    \caption{The structure of Higgs plus multi-gluon plus two quark-antiquark pair amplitudes obtained by
combining the MHV tower for $\phi+q\bar q+Q\bar Q +n$~gluons  and the anti-MHV tower
of $\phi^\dagger +q\bar q+Q\bar Q+n$~gluon amplitudes.}
    \label{fig:4qtower}
\end{figure}

The case of two quark-antiquark pairs proceeds in an almost identical way.
Helicity conservation
along both quark lines immediately leads us to the fact that the first non-zero $\phi$-amplitude
contains two negative helicities,
$A(q^{\lambda_1},\bar{q}^{-\lambda_1},Q^{\lambda_2},\bar{Q}^{-\lambda_2})$,
and this is precisely the third $\phi$-MHV amplitude in \eqn{phimhvbamps}.
Figure \ref{fig:4qtower}
shows the structure of the MHV and \MHVb~towers.
In the same way as before we add together the MHV(red)
and \MHVb(green) amplitudes to get the Higgs amplitudes.

In all three figures we combine the MHV amplitudes at tree level by first continuing
appropriate lines off-shell in the same way as in \cite{CSW1}, and then connecting
them by scalar propagators. The propagators connecting gluon or fermion lines
are always of the scalar type, $1/q^2$, as explained in \cite{GK,GGK}.

\section{Amplitudes with one quark-antiquark pair \label{sec:2q}}

When there is a single quark-antiquark pair, the tree-level amplitude
can be decomposed into colour-ordered amplitudes as follows,
\begin{eqnarray}
\lefteqn{
{\cal A}_n(\phi,\{p_i,\lambda_i,a_i\},\{p_j,\lambda_j,i_j\}) }\\
&&= 
i C g^{n-2}
\sum_{\sigma \in S_{n-2}} (T^{a_{\sigma(2)}}\cdots T^{a_{\sigma(n-1)}})_{i_1i_n}\,
A_n(\phi,1^{\lambda},\sigma(2^{\lambda_2},\ldots,{(n-1)}^{\lambda_{n-1}}),
n^{-\lambda})\,.\nonumber 
\label{TreeColorDecomposition}
\end{eqnarray}
where $S_{n-2}$ is the set of permutations of $(n-2)$ gluons.
Gluons are characterised with adjoint
colour label $a_i$, momentum $p_i$ and helicity $\lambda_i$ for
$i=1,\ldots,n-1$,
while the fermions carry fundamental colour label $i_j$, momentum $p_j$ and
helicity $\lambda_j$ for $j=1,n$.
By current conservation, the quark and antiquark helicities are  related such
that $\lambda_1 = -\lambda_n \equiv \lambda$ where $\lambda = \pm {1\over 2}$.

\subsection{MHV Amplitudes}
\begin{figure}[h]
\label{fig:mhv2q}
	\centering
	\psfrag{1}{$1^{\lambda}$}
	\psfrag{r-}{$r^-$}
	\psfrag{n}{$n^{-\lambda}$}
	\psfrag{phi}{$\phi$}
	\begin{center}
		\includegraphics[width=5cm]{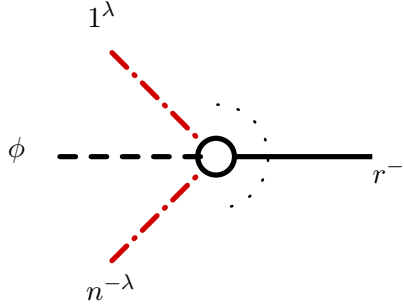}
	\end{center}
	\caption{MHV vertices for $\phi$ and one quark pair. The quark line is represented by
	the red dot-dashed line where $\pm\lambda_1$ are the helicities of the quark and anti-quark. 
	The negative helicity gluon is represented as a solid black line.}
	\label{fig:2Qmhv}
\end{figure}
There are two MHV vertices for a scalar, $\phi$, and a quark pair indicated in figure~\ref{fig:mhv2q}. 
Together with the usual MHV vertices~(\ref{mhvamps}) and a scalar with gluons~(\ref{assrtn}) 
we can begin to construct tree-level Higgs amplitudes with one quark pair. The expressions for the new vertices are:
\begin{eqnarray}
	A_n(\phi,q^{-}_1,\ldots,g^-_r,\ldots,\bar{q}^{+}_n) 
	&=& \frac{\spa{r}.{1}^3\spa{r}.{n}}{\prod_{l=1}^n\spa{l}.{l+1}} \label{eq:2qMHV1},\\
	A_n(\phi,q^{+}_1,\ldots,g^-_r,\ldots,\bar{q}^{-}_n) &=& 
	\frac{\spa{r}.{1}\spa{r}.{n}^3}{\prod_{l=1}^n\spa{l}.{l+1}}.
	\label{eq:2qMHV2}
\end{eqnarray}
The \MHVb~amplitudes can be obtained by the following parity transformation:
\begin{equation}
	A_n(\phi^\dagger,q^{\lambda}_1,\ldots,g^+_r,\ldots,\bar{q}^{-\lambda}_n) = (-1)^n\left(A_n(\phi,q^{-\lambda}_1,\ldots,g^-_r,\ldots,\bar{q}^{\lambda}_n)\right)^*
\end{equation}
Besides having the correct collinear and multi-particle factorization 
behavior, these amplitudes also correctly reduce to pure QCD 
amplitudes as the $\phi$ momentum approaches zero.
As discussed in Sec.~2, eqs.~(\ref{eq:2qMHV1}) and ~(\ref{eq:2qMHV2}) follow from
the analagous MHV amplitudes
for $\phi-$gluon interactions~(\ref{assrtn}) by supersymmetry.
Alternatively, eqs. (\ref{eq:2qMHV1}) and (\ref{eq:2qMHV2}) can be proved  recursively,
along the lines of the proof in the QCD case~\cite{BerendsGiele},
or using the light-cone recursive currents of
ref.~\cite{LightConeRecursive}.

\subsubsection{$H \to q^-g^-g^+\bar{q}^+$}
This amplitude corresponds to $n_++n_- = 4$, $n_+-n_-=0$. As we see from figure~\ref{fig:treemap}, the amplitude receives
contributions from both the MHV and $\overline{\text{MHV}}$ towers, so that
\begin{eqnarray}
	A_n(H,q^-_1,g^-_2,g^+_3,\bar{q}^+_4) &=& A_n(\phi,q^-_1,g^-_2,g^+_3,\bar{q}^+_4) + A_n(\phi^\dagger,q^-_1,g^-_2,g^+_3,\bar{q}^+_4) \nonumber\\
	&=& -\frac{[34]^2[13]}{[12][23][41]} - \frac{\spa{1}.{2}^2\spa{2}.{4}}{\spa{2}.{3}\spa{3}.{4}\spa{4}.{1}}.
\end{eqnarray}
This expression agrees with the known analytic formulae of Ref.~\cite{KDR}.
\subsection{NMHV Amplitudes}
\begin{figure}[t]
	\psfrag{j}{\small $j$}
	\psfrag{j+1}{\small $j+1$}
	\psfrag{i}{\small $i$}
	\psfrag{i+1}{\small $i+1$}
	\begin{center}
		\includegraphics[width=5cm]{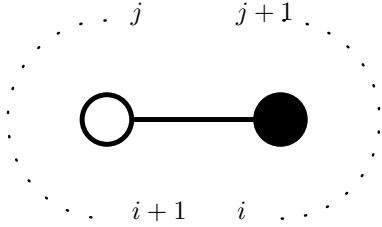}
	\end{center}
	\caption{Skeleton diagram showing the labelling of the positive helicity gluons in NMHV amplitudes. The gluons are shown as dotted lines with labels showing the bounding $g^+$ lines in each MHV vertex.}
	\label{fig:skel}
\end{figure}

\begin{figure}[t]
\psfrag{n+}{\Huge$n^{-\lambda}$}
\psfrag{1-}{\Huge$1^{\lambda}$}
\psfrag{phi}{\Huge$\phi$}
\psfrag{m2-}{\Huge$m_2^-$}
\psfrag{m3-}{\Huge$m_3^-$}
\psfrag{+}{\Huge$+$}
\psfrag{-}{\Huge$-$}
\begin{center}
{\scalebox{0.35}{
\includegraphics{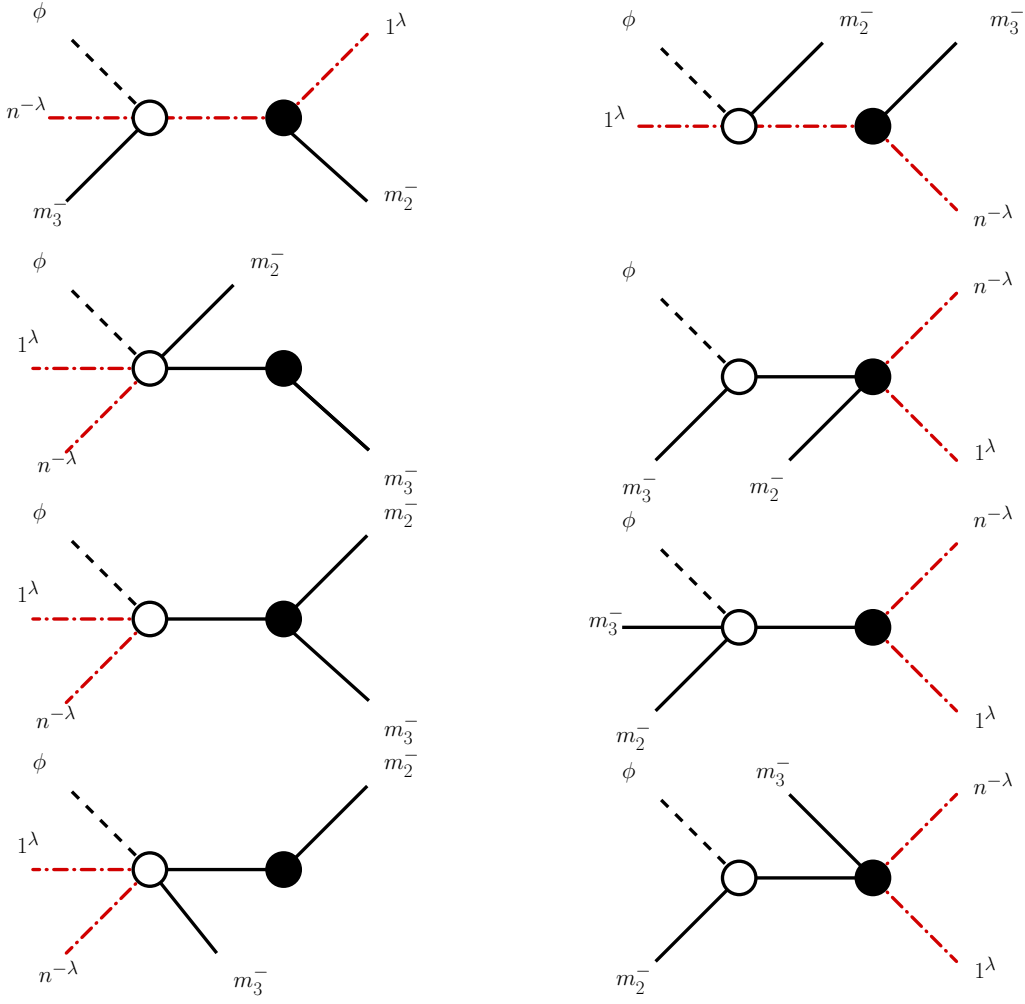}}
}
\end{center}
\caption{Tree diagrams with MHV vertices contributing to the
amplitude
$A_n(\phi,q_1^{\lambda},\ldots, m_2^-,\ldots,m_3^-,\ldots,\bar{q}_n^{-\lambda})$.
The scalar $\phi$ is represented by a dashed line and
negative helicity gluons, $g^-,$  by solid lines. The quark-antiquark line
is represented by the red dot-dashed line. }
\label{fig:nmhv2q}
\end{figure}

We continue by deriving the Next-to-MHV
(NMHV) amplitude 
\begin{equation}
\label{eq:2qNMHV}
	A_n(\phi,q_1^{\lambda},\ldots, m_2^-,\ldots,m_3^-,\ldots,\bar{q}_n^{-\lambda}),
\end{equation}
with three negative helicity particles -- one negative helicity quark(antiquark) and two negative
helicity gluons labelled as $m_2^-$ and $m_3^-$.
From now on we will suppress the dots for positive helicity gluons in the
MHV tower of amplitudes.
When labelling the partons in each NMHV diagram we systematically choose to put the $\phi$-MHV vertex
on the left. Figure \ref{fig:skel} shows a skeleton diagram of a generic NMHV amplitude and
shows how the partons are labelled cyclicly. The dotted semicircles denote the emission of positive helicity gluons 
from the vertex.
We use this convention in all of the NMHV diagrams with
one or two quark pairs.
All possible  diagrams contributing to (\ref{eq:2qNMHV}) are shown in 
figure~\ref{fig:nmhv2q}. 
Each of these diagrams is drawn for the fixed arrangement of negative helicity
gluons, such that
$q_1^{\lambda}$ is followed by $m_2^-$, followed by $m_3^-$ followed by
$\bar{q}_n^{-\lambda}$.
The full NMHV amplitude is given by,
\begin{equation}
\label{eq:NMHV}
A_n(\phi, 1^{\lambda}, m_2^-,m_3^-,n^{-\lambda}) = \, \frac{1}{\prod_{l=1}^n \spa{l}.{l+1}}\,
\sum_{i=1}^8 
A_n^{(i),\lambda} (m_2,m_3) \ ,
\end{equation}
where the common standard denominator of cyclic products of $\spa{l}.{l+1}$
is factored out for convenience.
We label the parton momenta as $p_i$ (where $i$ is defined modulo 
$n$)
and
introduce the composite (off-shell) momentum,
\begin{equation}
q_{i+1,j}=p_{i+1}+\ldots+\ldots+p_j.
\end{equation}
Note that the momentum of $\phi$, $p_\phi$, does not
enter the sum. In particular, $q_{i+1,i}=-p_{\phi}.$
As usual, the  off-shell continuation of the helicity spinor is defined as
\cite{CSW1},
\begin{equation}
\lambda_{i+1,j~\alpha} = q_{i+1,j \alpha \dot\alpha} \,\xi^{\dot\alpha},
\end{equation}
where $\xi^{\dot\alpha}$ is a reference spinor that can be arbitrarily chosen.
Following the organisational structure of \cite{GGK,DGK},
the contributions of the individual diagrams in figure~\ref{fig:nmhv2q} are given by
\begin{eqnarray}
A_n^{(1),\lambda}(m_2,m_3) &=& 
\sum_{i=m_2}^{m_3-1} \sum_{j=n}^{n}
\frac{ 
A_n^{(1),\lambda}(q_{j+1,i},m_2,m_3)
}
{D(j,i,q_{j+1,i})},\nonumber \\
A_n^{(3),\lambda}(m_2,m_3) &=& 
\sum_{i=m_3}^{n-1}\sum_{j=m_2}^{m_3-1} 
\frac{ 
A_n^{(3),\lambda}(q_{j+1,i},m_2,m_3)
}
{D(j,i,q_{j+1,i})},\nonumber \\
A_n^{(5),\lambda}(m_2,m_3) &=& 
\sum_{i=m_3}^{n-1}\sum_{j=1}^{m_2-1} 
\frac{ 
A_n^{(5),\lambda}(q_{j+1,i},m_2,m_3)
}
{D(j,i,q_{j+1,i})},\nonumber \\
A_n^{(7),\lambda}(m_2,m_3) &=& 
\sum_{i=m_2}^{m_3-1}\sum_{j=1}^{m_2-1} 
\frac{ 
A_n^{(7),\lambda}(q_{j+1,i},m_2,m_3)
}
{D(j,i,q_{j+1,i})},\nonumber \\
A_n^{(2k),\lambda}(q_{j+1,i},m_2,m_3) &=& 
\frac{
A_n^{(2k-1),\lambda}(q_{i+1,j},m_2,m_3)
}
{D(i,j,q_{i+1,j})}\qquad\text{for}\qquad k=1,\ldots,4\nonumber \\
\end{eqnarray}
and where
\begin{equation}
\label{eq:Ddef}
D(i,j,q) =
\langle i^- | \slash\!\!\! q |\xi^-\rangle
\langle j+1^- | \slash\!\!\! q |\xi^-\rangle
\langle i+1^- | \slash\!\!\! q |\xi^-\rangle
\langle j^- | \slash\!\!\! q |\xi^-\rangle
\frac{q^2}{\spa{i}.{i+1}\spa{j}.{j+1}}.
\end{equation}
The amplitudes where the quark carries negative helicity are given by:
\begin{eqnarray}
	A_n^{(1),-}(q,m_2,m_3) &=&
	\spa{m_2}.{1}^3 
	\spab{m_2}.{q}.{\xi}	
	\spab{m_3}.{q}.{\xi}^3
	\spa{m_3}.{n},
	\nonumber\\
	A_n^{(3),-}(q,m_2,m_3) &=&
	\spa{m_2}.{1}^3 
	\spa{m_2}.{n}	
	\spab{m_3}.{q}.{\xi}^4,
	\nonumber\\
	A_n^{(5),-}(q,m_2,m_3) &=&
	\spab{1}.{q}.{\xi}^3 
	\spab{n}.{q}.{\xi}
	\spa{m_2}.{m_3}^4,
	\nonumber\\
	A_n^{(7),-}(q,m_2,m_3) &=&
	\spa{m_3}.{1}^3 
	\spa{m_3}.{n}
	\spab{m_2}.{q}.{\xi}^4,
		\label{eq:2q-}
\end{eqnarray}
while the amplitudes where the quark carries positive helicity are given by,
\begin{eqnarray}
A_n^{(1),+}(q,m_2,m_3) &=& 
\spa{m_2}.{1} 
\spab{m_2}.{q}.{\xi}^3
\spab{m_3}.{q}.{\xi}
\spa{m_3}.{n}^3,
\nonumber \\
A_n^{(3),+}(q,m_2,m_3) &=& 
\spa{m_2}.{1} 
\spa{m_2}.{n}^3
\spab{m_3}.{q}.{\xi}^4,
\nonumber \\
A_n^{(5),+}(q,m_2,m_3) &=& 
\spab{1}.{q}.{\xi}
\spab{n}.{q}.{\xi}^3
\spa{m_2}.{m_3}^4,
\nonumber \\
A_n^{(7),+}(q,m_2,m_3) &=& 
\spa{m_3}.{1} 
\spa{m_3}.{n}^3
\spab{m_2}.{q}.{\xi}^4.
\label{eq:2q+}
\end{eqnarray}

As in Ref.~\cite{DGK} we leave the reference spinor $\xi$  
arbitrary and specifically do not set it to be
equal to one of the momenta in the problem.
This has two advantages.   First, we do not introduce
unphysical singularities in diagrams containing  a three gluon vertex.
Second, it  allows a powerful numerical check of gauge
invariance i.e. all colour ordered amplitudes must be independent 
of the specific choice of $\xi$.

Equation~\ref{eq:NMHV} describes all amplitudes
coupling $\phi$ to a quark-antiquark pair, 
2 negative helicity gluons and
any number of positive helicity gluons.
In particular, it describes $\phi \to q^{-}g^{-}g^{-}\bar{q}^{+}$. 
This final state only receives contributions
from the MHV tower of amplitudes and the amplitude
for $\phi \to q^{-}g^{-}g^{-}\bar{q}^{+}$ is therefore equivalent to
the amplitude for
$H \to q^{-}g^{-}g^{-}\bar{q}^{+}$.

From the amplitudes (\ref{eq:2q-}) and (\ref{eq:2q+}) we can observe that in the limit $p_\phi \to
0$ each even numbered diagram collapses on to the corresponding odd numbered diagram. The momentum
conservation law $q_{1,n} = p_\phi\to 0 $ implies that $q_{i+1,j}=-q_{j+1,i}$ i.e. the
transformation $i\leftrightarrow j$ leaves the amplitude unchanged as there are even numbers of
$q$'s in the expressions. This means that we recover the 4 NMHV quark-gluon diagrams twice.


\subsubsection{$H \to q^{-}g^{-}g^{-}\bar{q}^{+}$}

In this case, we can take $\lambda = -$, $m_2 = 2$ and $m_3=3$.  
The third and seventh classes of diagrams in figure~\ref{fig:nmhv2q}
 collapse since there are
not enough gluons to prevent the right hand vertex vanishing. 

We have checked, 
with a help of a symbolic manipulator, that our results are $\xi$-independent
(gauge invariant) and  numerically agree with the known analytic formulae~\cite{KDR},
\begin{eqnarray}
A_4(H,1_q^{-},2^{-},3^{-},4_{\bar{q}}^{+})&=& 
\frac{\langle 3^-|\slash\!\!\! p_H | 4^- \rangle^2 
\spa{1}.{2}}{\spb{4}.{2} s_{124}}\left(\frac{1}{s_{12}}+\frac{1}{s_{14}}\right)
-\frac{\langle 2^-|\slash\!\!\! p_H | 4^- \rangle^2 \spa{1}.{3}}{\spb{4}.{3}s_{134} s_{14}}
\nonumber \\
&&-\frac{\langle 1^-|\slash\!\!\! p_H | 4^- \rangle^2 }
{\spa{1}.{4}\spb{4}.{2}\spb{4}.{3}\spb{2}.{3}}
\end{eqnarray}
where $p_H = p_\phi$.


\subsubsection{$H \to q^{\lambda}g^{-}g^{-}g^{+}\bar{q}^{-\lambda}$}

As discussed in Ref.~\cite{DFM}, there are three independent amplitudes, 
corresponding to having any of the three gluons 
with positive helicity. Each amplitude receives contributions from both the MHV and anti-MHV 
towers so that setting $(m_2,m_3)$ to be $(2,3)$, for example,
\begin{eqnarray}
\label{eq:5ptest}
A_5(H,1_q^\lambda,2^-,3^-,4^+,5_{\bar q}^{-\lambda}) 
&=& A_5(\phi,1_q^\lambda,2^-,3^-,4^+,5_{\bar q}^{-\lambda}) \nonumber\\
&+& A_5(\phi^\dagger,1_q^\lambda,2^-,3^-,4^+,5_{\bar q}^{-\lambda}) . 
\end{eqnarray}
We can obtain the negative helicites in other positions by taking 
$(m_2,m_3)$ to be either  $(2,4)$ or $(3,4)$.
We have checked numerically that eq.~(\ref{eq:5ptest}) is gauge invariant and gives the 
same result as an independent Feynman diagram calculation.   The same holds for the other assignments 
of negative helicity gluons.

\newpage
\section{Amplitudes with two quark-antiquark pairs \label{sec:4q}}

When there are two quark-antiquark pairs the 
tree-level amplitude can be decomposed into colour ordered 
amplitudes as,
\begin{eqnarray}
	&&\mathcal{A}_n(\phi,\{p_i,\lambda_i,a_i\},\{p_j,\lambda_j,i_j\}) 
	= iC^\prime g^{n-2} \sum_k^{n-4}\sum_{\sigma\in S_k}\sum_{\rho\in S_l} \bigg\{ \nonumber\\
	&\phantom{-}&(T^{a_{\sigma(1)}}\cdots T^{a_{\sigma(k)}})_{i_1 i_4} (T^{a_{\rho(1)}}\cdots T^{a_{\rho(l)}})_{i_3 i_2} \nonumber\\
	&&\times
	A_n(\phi,q_1^{\lambda_1},\sigma(1),\ldots,\sigma(k)),\bar{Q}^{-\lambda_2}_s;
	Q^{\lambda_2}_{s+1},\rho(1),\ldots,\rho(l),\bar{q}^{-\lambda_1}_n) \nonumber\\
	&-&\frac{1}{N}(T^{a_{\sigma(1)}}\cdots T^{a_{\sigma(k)}})_{i_1 i_2} (T^{a_{\rho(1)}}\cdots T^{a_{\rho(l)}})_{i_3 i_4} \nonumber\\
	&&\times\tilde{A}_n(\phi,q_1^{\lambda_1},\sigma(1),\ldots,\sigma(k),\bar{q}^{-\lambda_1}_s;
	Q^{\lambda_2}_{s+1},\rho(1),\ldots,\rho(l),\bar{Q}^{-\lambda_1}_n)\bigg\}
	\label{eq:qqQQcoldecomp}
\end{eqnarray}
where $S_k$ and $S_l$ are permutation groups such that $k+l=n-4$ and represent the possible ways of distributing the gluons in a colour ordered way between
the quarks. For $i=j=0$, $(T^{a_i}\ldots T^{a_j})_{kl}$ reduces to $\delta_{kl}$.  The first quark-antiquark pair have fundamental colour indices $i_1$ and
$i_2$ respectively with  helicities $\lambda_1$,$-\lambda_1$ whereas the second quark-antiquark pair have fundamental colour indices $i_3$ and $i_4$ with
helicities $\lambda_2$,$-\lambda_2$. We see that the two amplitudes $A_n$ and $\tilde{A}_n$ correspond to different ways of connecting the fundamental
colour charges.
For the $A$ amplitudes, there is a colour line connecting $q$ and $\bar Q$ and a second line connecting $Q$ and $\bar q$, while for 
the QED-like $\tilde A$ amplitudes the colour lines connect $q$ to $\bar q$ and $Q$ to $\bar Q$. Any number of gluons may be radiated from each colour
line.

\subsection{MHV Amplitudes}
\begin{figure}[h]
	\centering
	\psfrag{1}{$1^{\lambda_1}$}
	\psfrag{s}{$s^{-\lambda_2}$}
	\psfrag{s2}{$s^{-\lambda_1}$}
	\psfrag{s+1}{$(s+1)^{\lambda_2}$}
	\psfrag{s2+1}{$(s+1)^{\lambda_2}$}
	\psfrag{n}{$n^{-\lambda_1}$}
	\psfrag{n2}{$n^{-\lambda_2}$}
	\psfrag{phi}{$\phi$}
	\psfrag{(a)}{(a)}
	\psfrag{(b)}{(b)}
	\begin{center}
		\includegraphics[width=10cm]{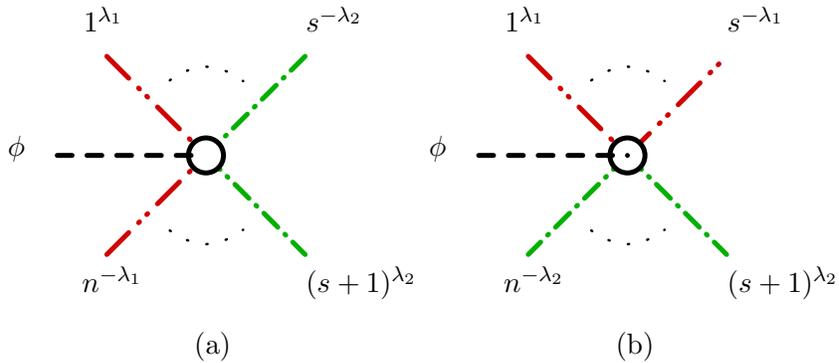}
	\end{center}
	\caption{MHV vertices for $\phi$ and two quark pairs. Quarks of the same flavour are represented by
	lines of the same type - i.e. red or green dot-dashed lines.  The colour connection is 
	organised cyclically and each colour connected quark-antiquark pair may have any number of 
	positive helicity gluons radiated from it, represented by the dots. 
	$\lambda_1$,$\lambda_2$ are the helicities of the quarks. Figure (a) represents the colour
	ordering for the `$A$' amplitudes while (b) represents the colour ordering for the
	`$\tilde{A}$' amplitudes.}
	\label{fig:4Qmhv}
\end{figure}
For each colour structure there are four MHV amplitudes where two of the fermions have 
negative helicity and two have positive helicity as shown in
figure~\ref{fig:4Qmhv}. 
Any number of positive helicity gluons can be radiated from each of the quark colour lines. 
Figure \ref{fig:4Qmhv} explicitly shows the two ways of connecting up the colours. For each 
helicity configuration we can write,
\begin{eqnarray}
\label{eq:4qmhv1}
	A_n(\phi,q^+_1,\ldots,\bar{Q}^-_s,Q^+_{s+1},\ldots,\bar{q}^-_n) &=& \frac{\spa{1}.{s}\spa{s}.{n}^2\spa{n}.{s+1}}{\prod_{l=1}^n \spa{l}.{l+1}} ,\\
\label{eq:4qmhv2}
	A_n(\phi,q^+_1,\ldots,\bar{Q}^+_s,Q^-_{s+1},\ldots,\bar{q}^-_n) &=& \frac{\spa{1}.{s}\spa{n}.{s+1}^3}{\prod_{l=1}^n \spa{l}.{l+1}} ,\\
\label{eq:4qmhv3}
	A_n(\phi,q^-_1,\ldots,\bar{Q}^+_s,Q^-_{s+1},\ldots,\bar{q}^+_n) &=& \frac{\spa{s}.{1}\spa{1}.{s+1}^2\spa{s+1}.{n}}{\prod_{l=1}^n \spa{l}.{l+1}} ,\\
\label{eq:4qmhv4}
	A_n(\phi,q^-_1,\ldots,\bar{Q}^-_s,Q^+_{s+1},\ldots,\bar{q}^+_n) &=& \frac{\spa{1}.{s}^3\spa{n}.{s+1}}{\prod_{l=1}^n \spa{l}.{l+1}},
\end{eqnarray}
with the other colour ordering given by,
\begin{eqnarray}
\label{eq:4qmhv5}
	\tilde{A}_n(\phi,q^+_1,\ldots,\bar{q}^-_s,Q^+_{s+1},\ldots,\bar{Q}^-_n) &=& \frac{\spa{1}.{n}\spa{n}.{s}^2\spa{s}.{s+1}}{\prod_{l=1}^n \spa{l}.{l+1}} ,\\
\label{eq:4qmhv6}
	\tilde{A}_n(\phi,q^+_1,\ldots,\bar{q}^-_s,Q^-_{s+1},\ldots,\bar{Q}^+_n) &=& \frac{\spa{1}.{n}\spa{s}.{s+1}^3}{\prod_{l=1}^n \spa{l}.{l+1}} ,\\
\label{eq:4qmhv7}
	\tilde{A}_n(\phi,q^-_1,\ldots,\bar{q}^+_s,Q^-_{s+1},\ldots,\bar{Q}^+_n) &=& \frac{\spa{n}.{1}\spa{1}.{s+1}^2\spa{s+1}.{s}}{\prod_{l=1}^n \spa{l}.{l+1}} ,\\
\label{eq:4qmhv8}
	\tilde{A}_n(\phi,q^-_1,\ldots,\bar{q}^+_s,Q^+_{s+1},\ldots,\bar{Q}^-_n) &=& \frac{\spa{1}.{n}^3\spa{s}.{s+1}}{\prod_{l=1}^n \spa{l}.{l+1}}.
	\label{eq:4qMHVamps2}
\end{eqnarray}
These MHV vertices are derived from 
the $\phi$-gluon vertices using a (pseudo) ${\cal N}=2$ supersymmetric Ward identity as discussed in section~2.
The  amplitudes involving $\phi^\dagger$ are related by parity and can be obtained by conjugating
the MHV expressions,
\begin{equation}
	A(\phi^\dagger,q_1^{\lambda_1},\bar{Q}^{-\lambda_2}_{s};Q^{\lambda_2}_{s+1},\bar{q}^{-\lambda_1}_n) =  (-1)^n\left(A(\phi,q_1^{-\lambda_1},\bar{Q}^{\lambda_2}_{s};Q^{-\lambda_2}_{s+1},\bar{q}^{\lambda_1}_n)\right)^*,
\end{equation}
and similarly for the $\tilde{A}$ amplitudes.

Equations~(\ref{eq:4qmhv1})--(\ref{eq:4qmhv8}) have
an identical form to the pure QCD amplitudes.
As such, they have the correct collinear and multi-particle factorization 
behavior and a correct limit as the $\phi$ momentum approaches zero.

\subsubsection{$H\to q^-\bar{Q}^+Q^-\bar{q}^+$}
When $n=4$, there is only one possibility, $n_+ = n_- = 2$. 
As can be seen from figure~\ref{fig:4qtower}, this lies in the intersection
of the MHV and $\overline{\rm MHV}$ towers so that, setting $s=2$,
\begin{eqnarray}
	A_4(H,q^-_1,\bar{Q}^+_2,Q^-_3,\bar{q}^+_4) &=& 
	A_4(\phi,q^-_1,\bar{Q}^+_2,Q^-_3,\bar{q}^+_4)
       +A_4(\phi^\dagger,q^-_1,\bar{Q}^+_2,Q^-_3,\bar{q}^+_4) \nonumber\\
	&=& -\frac{\spa{1}.{3}^2}{\spa{2}.{3}\spa{4}.{1}} - \frac{[24]^2}{[23][41]}
\end{eqnarray}
which agrees with the known analytic formulae of ref.~\cite{KDR}.
\subsection{NMHV Amplitudes}
There are four different helicity configurations for amplitudes with two quark pairs and a single negative helicity gluon.
We choose the first quark pair to have helicities $\pm\lambda_1$ and the second pair to carry helicities $\pm\lambda_2$. 
Again suppressing the positive helicity gluons, we can write the NMHV amplitude as,
\begin{eqnarray}
\label{eq:4qnmhv1}
	A_{n}(\phi,q_1^{\lambda_1},g_{m_2}^-,\bar{Q}^{-\lambda_2}_{m_3};Q^{\lambda_2}_{m_3+1},\bar{q}^{-\lambda_1}_n) = 
	\frac{1}{\prod_{l=1}^n \spa{l}.{l+1}}\sum_{i=1}^{10} A^{(i),\lambda_1\lambda_2}_{1:n}(m_2,m_3) ,\\
\label{eq:4qnmhv2}
	\tilde{A}_{n}(\phi,q_1^{\lambda_1},g_{m_2}^-,\bar{q}^{-\lambda_1}_{m_3};Q^{\lambda_2}_{m_3+1},\bar{Q}^{-\lambda_2}_n) = 
	\frac{1}{\prod_{l=1}^n \spa{l}.{l+1}}\sum_{i=1}^{8} \tilde{A}^{(i),\lambda_1\lambda_2}_{1:n}(m_2,m_3).
\end{eqnarray}
There are two other amplitudes where the negative helicity gluon appears on the other quark line, for example,
$$
        A_{n}(\phi,q_1^{\lambda_1},\bar{Q}^{-\lambda_2}_{m_3};Q^{\lambda_2}_{m_3+1},g_{m_2}^-,\bar{q}^{-\lambda_1}_n),
$$
 however these amplitudes can be obtained by using the property that the amplitudes are cyclic in the quark lines,
 we can move the gluon from one quark colour line to the other
 by exchanging the two lines and relabelling,
 $q_1 \leftrightarrow  Q^{\lambda_2}_{m_3+1}$, $\bar{Q}^{-\lambda_2}_{m_3} \leftrightarrow \bar{q}^{-\lambda_1}_n$.
A similar relabelling applies to $\tilde{A}_n$
\begin{figure}[h!]
	\centering
	\psfrag{m2-}{\scriptsize$m_2^-$}
	\psfrag{1-}{\scriptsize$1^{\lambda_1}$}
	\psfrag{m3+1-}{\scriptsize$(m_3+1)^{\lambda_2}$}
	\psfrag{m3+}{\scriptsize$m_3^{-\lambda_2}$}
	\psfrag{phi}{\scriptsize$\phi$}
	\psfrag{n+}{\scriptsize$n^{-\lambda_1}$}
	\includegraphics[width=0.90\textwidth]{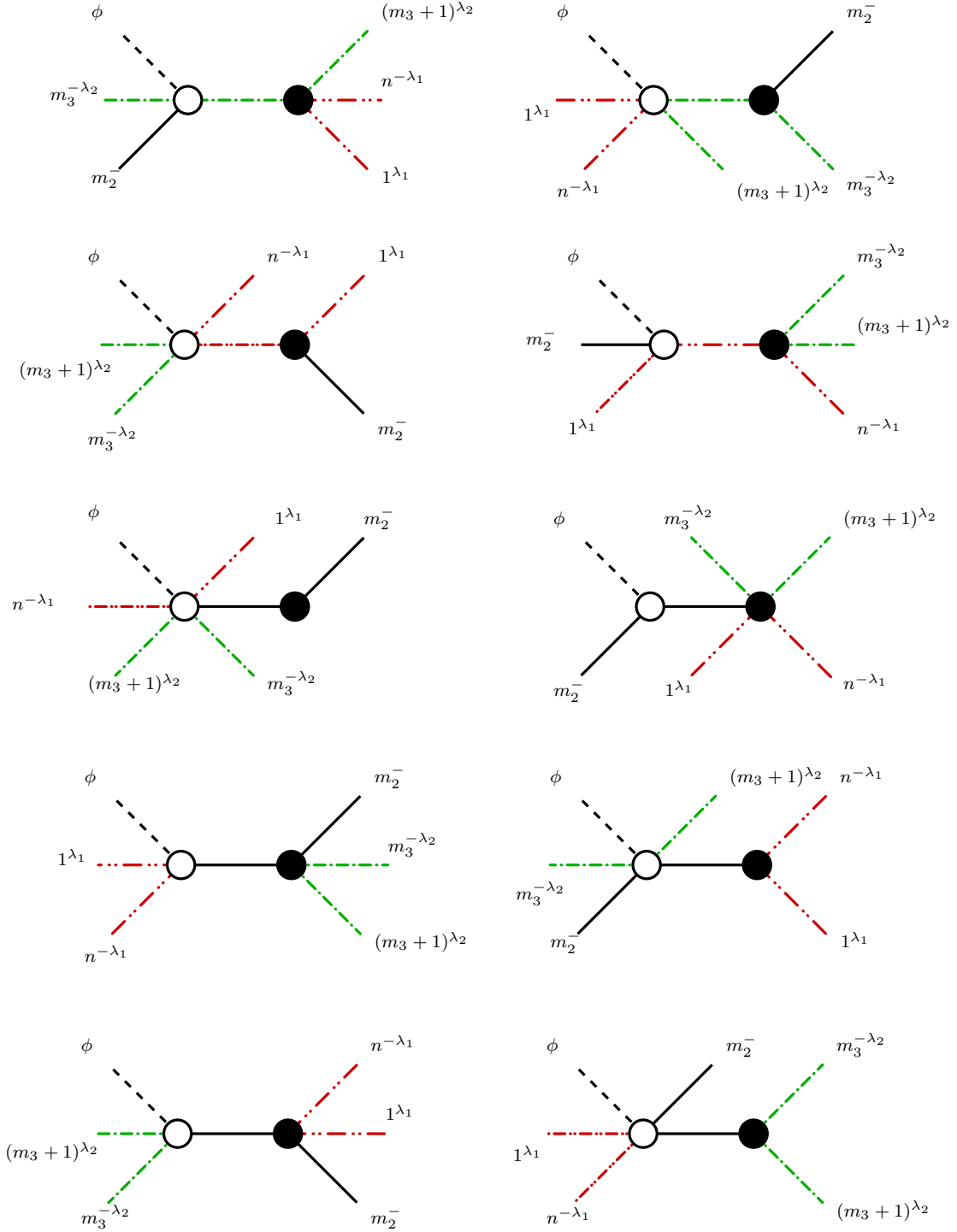}
	\caption{Tree diagrams with MHV vertices contributing to the amplitude
	$A_{n}(\phi,q_1^{\lambda_1},g_{m_2}^-,\bar{Q}^{-\lambda_2}_{m_3};Q^{\lambda_2}_{m_3+1},\bar{q}^{-\lambda_1}_n)$.
	The scalar $\phi$ is represented by a dashed line whereas the two quark lines a represented
	by coloured dot-dashed lines. The negative helicity gluons are the solid black lines.}
	\label{fig:nmhv4Qi}
\end{figure}

The 10 diagrams describing the $A_n$ colour ordering are shown in figure~\ref{fig:nmhv4Qi}.
The resulting amplitudes are given by,
\begin{eqnarray}
	A_{n}^{(1),\lambda_1\lambda_2}(m_2,m_3) &=& 
	\sum_{i=1}^{m_2-1}\sum_{j=m_3}^{m_3} 
	\frac{ 
	A_n^{(1),\lambda_1\lambda_2}(q_{j+1,i};m_2,m_3)
	}
	{D(j,i,q_{j+1,i})} ,\nonumber \\
	A_{n}^{(3),\lambda_1\lambda_2}(m_2,m_3) &=& 
	\sum_{i=m_2}^{m_3-1}\sum_{j=n}^{n} 
	\frac{
	A_n^{(3),\lambda_1\lambda_2}(q_{j+1,i};m_2,m_3)
	}
	{D(j,i,q_{j+1,i})} ,\nonumber \\
	A_{n}^{(5),\lambda_1\lambda_2}(m_2,m_3) &=& 
	\sum_{i=m_2}^{m_3-1}\sum_{j=1}^{m_2-1} 
	\frac{ 
	A_n^{(5),\lambda_1\lambda_2}(q_{j+1,i};m_2,m_3)
	}
	{D(j,i,q_{j+1,i})} ,\nonumber \\
	A_{n}^{(7),\lambda_1\lambda_2}(m_2,m_3) &=& 
	\sum_{i=m_3+1}^{n-1}\sum_{j=1}^{m_2-1} 
	\frac{
	A_n^{(7),\lambda_1\lambda_2}(q_{j+1,i};m_2,m_3)
	}
	{D(j,i,q_{j+1,i})} ,\nonumber \\
	A_{n}^{(9),\lambda_1\lambda_2}(m_2,m_3) &=& 
	\sum_{i=m_2}^{m_3-1}\sum_{j=m_3+1}^{n-1} 
	\frac{
	A_n^{(9),\lambda_1\lambda_2}(q_{j+1,i};m_2,m_3)
	}
	{D(j,i,q_{j+1,i})} ,
\end{eqnarray}
and	
\begin{eqnarray}
A_{n}^{(2k),\lambda_1\lambda_2}(q_{j+1,i};m_2,m_3) &=& \frac{A_{n}^{(2k-1),\lambda_1\lambda_2}(q_{i+1,j};m_2,m_3)}{D(i,j,q_{i+1,j})}
\qquad\text{for }\qquad\ k=1,\ldots,5.\nonumber \\ 
	\label{eq:4qamps1}
\end{eqnarray}
The quantity $D(i,j,q)$ is defined as in equation (\ref{eq:Ddef}). The amplitudes for each helicity configuration are given by:
\begin{eqnarray}
	A^{(1),--}(q;m_2,m_3) &=& 
	\spab{m_2}.{q}.{\xi}^3 
	\spa{m_2}.{m_3}
	\spa{n}.{m_3+1} 
	\spa{m_3+1}.{1}^2
	\spab{1}.{q}.{\xi}
	,\nonumber\\
	A^{(3),--}(q;m_2,m_3) &=& 
	\spa{m_2}.{1}^3
	\spab{m_2}.{q}.{\xi}
	\spa{m_3+1}.{n}
	\spab{m_3+1}.{q}.{\xi}^2
	\spab{m_3}.{q}.{\xi}
	,\nonumber\\
	A^{(5),--}(q;m_2,m_3) &=& 
	\spa{m_3}.{1}
	\spa{1}.{m_3+1}^2 
	\spa{m_3+1}.{n}
	\spab{m_2}.{q}.{\xi}^4
	,\nonumber\\
	A^{(7),--}(q;m_2,m_3) &=& 
	\spab{1}.{q}.{\xi}^3
	\spab{n}.{q}.{\xi}
	\spa{m_2}.{m_3+1}^3
	\spa{m_2}.{m_3}	
	,\nonumber\\
	A^{(9),--}(q;m_2,m_3) &=& 
	\spab{m_3+1}.{q}.{\xi}^3
	\spab{m_3}.{q}.{\xi}
	\spa{m_2}.{1}^3
	\spa{m_2}.{n},	
\\
	A^{(1),-+}(q;m_2,m_3) &=& 
	\spab{m_2}.{q}.{\xi} 
	\spa{m_2}.{m_3}^3
	\spa{n}.{m_3+1}
	\spab{1}.{q}.{\xi}^3
	,\nonumber\\
	A^{(3),-+}(q;m_2,m_3) &=& 
	\spa{m_2}.{1}^3
	\spab{m_2}.{q}.{\xi}
	\spab{m_3}.{q}.{\xi}^3
	\spa{m_3+1}.{n}
	,\nonumber\\
	A^{(5),-+}(q;m_2,m_3) &=& 
	\spa{m_3}.{1}^3 
	\spa{m_3+1}.{n}
	\spab{m_2}.{q}.{\xi}^4
	,\nonumber\\
	A^{(7),-+}(q;m_2,m_3) &=& 
	\spab{1}.{q}.{\xi}^3
	\spab{n}.{q}.{\xi}
	\spa{m_2}.{m_3+1}
	\spa{m_2}.{m_3}^3	
	,\nonumber\\
	A^{(9),-+}(q;m_2,m_3) &=& 
	\spab{m_3+1}.{q}.{\xi}
	\spab{m_3}.{q}.{\xi}^3
	\spa{m_2}.{1}^3
	\spa{m_2}.{n}	
,\\
	A^{(1),+-}(q;m_2,m_3) &=& 
	\spab{m_2}.{q}.{\xi}^3 
	\spa{m_2}.{m_3}
	\spa{n}.{m_3+1}^3
	\spab{1}.{q}.{\xi}
	,\nonumber\\
	A^{(3),+-}(q;m_2,m_3) &=& 
	\spa{m_2}.{1}
	\spab{m_2}.{q}.{\xi}^3
	\spab{m_3}.{q}.{\xi}
	\spa{m_3+1}.{n}^3
	,\nonumber\\
	A^{(5),+-}(q;m_2,m_3) &=& 
	\spa{m_3}.{1}
	\spa{m_3+1}.{n}^3
	\spab{m_2}.{q}.{\xi}^4
	,\nonumber\\
	A^{(7),+-}(q;m_2,m_3) &=& 
	\spab{1}.{q}.{\xi}
	\spab{n}.{q}.{\xi}^3
	\spa{m_2}.{m_3+1}^3
	\spa{m_2}.{m_3}	
	,\nonumber\\
	A^{(9),+-}(q;m_2,m_3) &=& 
	\spab{m_3+1}.{q}.{\xi}^3
	\spab{m_3}.{q}.{\xi}
	\spa{m_2}.{1}
	\spa{m_2}.{n}^3	
,\\
	A^{(1),++}(q;m_2,m_3) &=& 
	\spab{m_2}.{q}.{\xi} 
	\spa{m_2}.{m_3}^3
	\spa{n}.{m_3+1}
	\spab{n}.{q}.{\xi}^2	
	\spab{1}.{q}.{\xi}
	,\nonumber\\
	A^{(3),++}(q;m_2,m_3) &=& 
	\spa{m_2}.{1}
	\spab{m_2}.{q}.{\xi}^3
	\spab{m_3}.{q}.{\xi}
	\spa{m_3}.{n}^2
	\spa{m_3+1}.{n}
	,\nonumber\\
	A^{(5),++}(q;m_2,m_3) &=& 
	\spa{1}.{m_3} 
	\spa{m_3}.{n}^2
	\spa{n}.{m_3+1}
	\spab{m_2}.{q}.{\xi}^4
	,\nonumber\\
	A^{(7),++}(q;m_2,m_3) &=& 
	\spab{1}.{q}.{\xi}
	\spab{n}.{q}.{\xi}^3
	\spa{m_2}.{m_3+1}
	\spa{m_2}.{m_3}	^3
	,\nonumber\\
	A^{(9),++}(q;m_2,m_3) &=& 
	\spab{m_3+1}.{q}.{\xi}
	\spab{m_3}.{q}.{\xi}^3
	\spa{m_2}.{1}
	\spa{m_2}.{n}^3	.
\end{eqnarray}
\begin{figure}[h!]
	\centering
	\psfrag{m2-}{\scriptsize$m_2^-$}
	\psfrag{1-}{\scriptsize$1^{\lambda_1}$}
	\psfrag{m3+1-}{\scriptsize$(m_3+1)^{\lambda_2}$}
	\psfrag{m3+}{\scriptsize$m_3^{-\lambda_2}$}
	\psfrag{phi}{\scriptsize$\phi$}
	\psfrag{n+}{\scriptsize$n^{-\lambda_1}$}
	\includegraphics[width=0.9\textwidth]{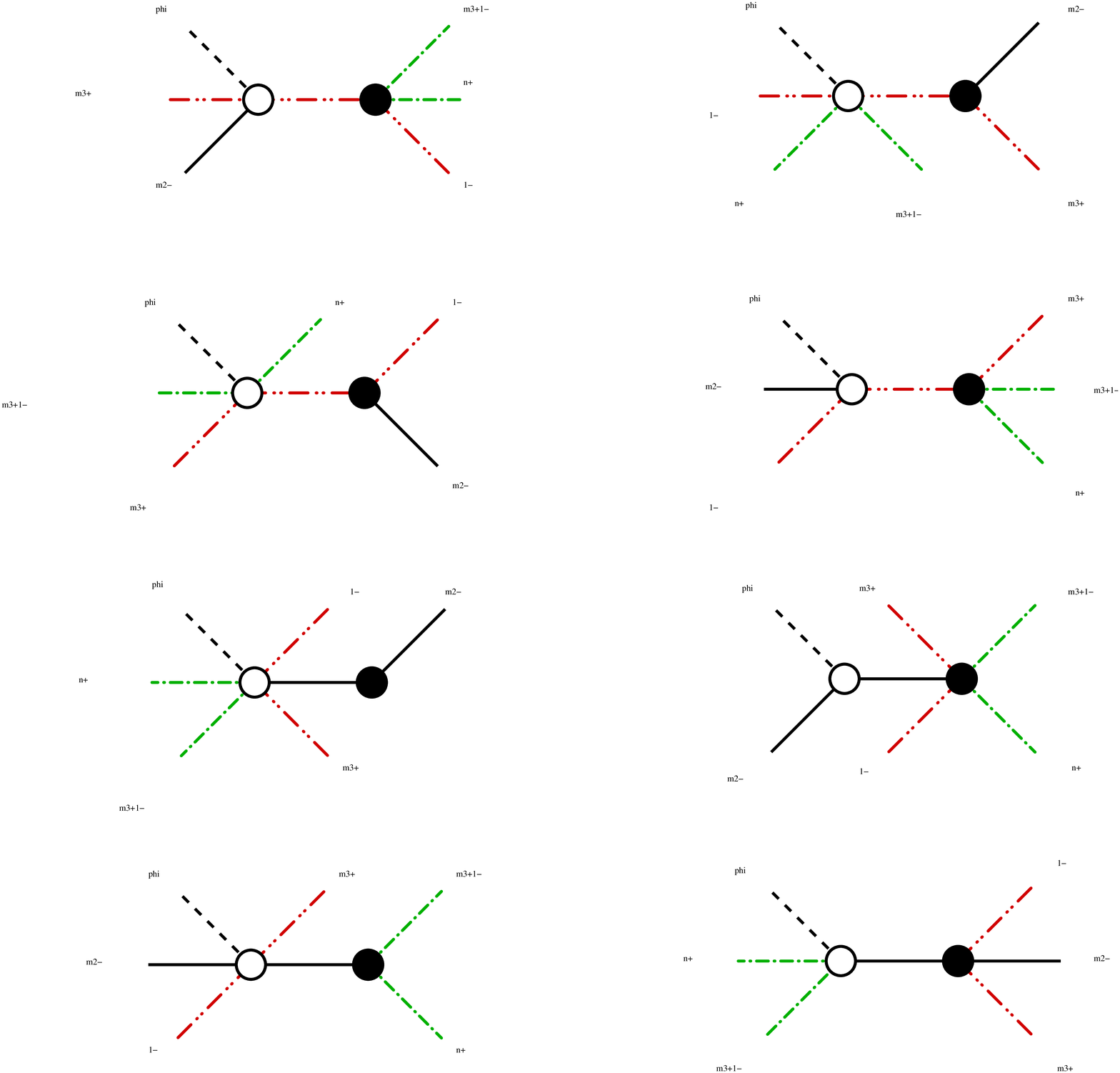}
	\caption{Tree diagrams with MHV vertices contributing to the amplitude
	$\tilde{A}_{n}(\phi,q_1^{\lambda_1},g_{m_2}^-,\bar{q}^{-\lambda_1}_{m_3};Q^{\lambda_2}_{m_3+1},\bar{Q}^{-\lambda_2}_n)$.
	The scalar $\phi$ is represented by a dashed line whereas the two quark lines a represented
	by coloured dot-dashed lines. The negative helicity gluons are the solid black lines.}
	\label{fig:nmhv4Qii}
\end{figure}
For the $\tilde{A}$ colour ordering there are only 8 diagrams shown in figure~\ref{fig:nmhv4Qii}. The corresponding
amplitudes are given by,
\begin{eqnarray}
	\tilde{A}_{n}^{(1),\lambda_1\lambda_2}(m_2,m_3) &=& 
	\sum_{i=1}^{m_2-1}\sum_{j=m_3}^{m_3} 
	\frac{
	\tilde{A}_n^{(1),\lambda_1\lambda_2}(q_{j+1,i};m_2,m_3)
	}
	{D(j,i,q_{j+1,i})} ,\nonumber \\
	\tilde{A}_{n}^{(3),\lambda_1\lambda_2}(m_2,m_3) &=& 
	\sum_{i=m_2}^{m_3-1}\sum_{j=n}^{n} 
	\frac{
	\tilde{A}_n^{(3),\lambda_1\lambda_2}(q_{j+1,i};m_2,m_3)
	}
	{D(j,i,q_{j+1,i})} ,\nonumber \\
	\tilde{A}_{n}^{(5),\lambda_1\lambda_2}(m_2,m_3) &=& 
	\sum_{i=m_2}^{m_3-1}\sum_{j=1}^{m_2-1} 
	\frac{
	\tilde{A}_n^{(5),\lambda_1\lambda_2}(q_{j+1,i};m_2,m_3)
	}
	{D(j,i,q_{j+1,i})},\nonumber \\
	\tilde{A}_{n}^{(7),\lambda_1\lambda_2}(m_2,m_3) &=& 
	\sum_{i=n}^{n}\sum_{j=m_3}^{m_3} 
	\frac{
	\tilde{A}_n^{(7),\lambda_1\lambda_2}(q_{j+1,i};m_2,m_3)
	}
	{D(j,i,q_{j+1,i})},
\end{eqnarray}
with
\begin{eqnarray}	
\tilde{A}_{n}^{(2k),\lambda_1\lambda_2}(q_{j+1,i};m_2,m_3) &=& \frac{\tilde{A}_{n}^{(2k-1),\lambda_1\lambda_2}(q_{i+1,j};m_2,m_3)}{D(i,j,q_{i+1,j})}
\qquad\text{for}\qquad\ k=1,\ldots,4\nonumber\\
	\label{eq:4qamps2}
\end{eqnarray}
The amplitudes for each helicity combination are,
\begin{eqnarray}
	\tilde{A}_n^{(1),--}(q;m_2,m_3) &=&
	\spa{m_2}.{m_3}
	\spab{m_2}.{q}.{\xi}^3
	\spa{n}.{1}
	\spa{1}.{m_3+1}^2
	\spab{m_3+1}.{q}.{\xi}
	,\nonumber\\
	\tilde{A}_n^{(3),--}(q;m_2,m_3) &=&
	\spab{m_2}.{q}.{\xi}
	\spa{m_2}.{1}^3
	\spab{n}.{q}.{\xi}
	\spab{m_3+1}.{q}.{\xi}^2
	\spa{m_3+1}.{m_3}
	,\nonumber\\
	\tilde{A}_n^{(5),--}(q;m_2,m_3) &=&
	\spa{n}.{1}
	\spa{1}.{m_3+1}^2
	\spa{m_3+1}.{m_3}
	\spab{m_2}.{q}.{\xi}^4	
	,\nonumber\\
	\tilde{A}_n^{(7),--}(q;m_2,m_3) &=&
	\spa{m_2}.{1}^3
	\spa{m_2}.{m_3}
	\spab{m_3+1}.{q}.{\xi}^3
	\spab{n}.{q}.{\xi},
\end{eqnarray}
\begin{eqnarray}
	\tilde{A}_n^{(1),-+}(q;m_2,m_3) &=&
	\spa{m_2}.{m_3}
	\spab{m_2}.{q}.{\xi}^3
	\spa{n}.{1}^3
	\spab{m_3+1}.{q}.{\xi}
	,\nonumber\\
	\tilde{A}_n^{(3),-+}(q;m_2,m_3) &=&
	\spab{m_2}.{q}.{\xi}
	\spa{m_2}.{1}^3
	\spab{n}.{q}.{\xi}^3
	\spa{m_3+1}.{m_3}
	,\nonumber\\
	\tilde{A}_n^{(5),-+}(q;m_2,m_3) &=&
	\spa{1}.{n}^3
	\spa{m_3}.{m_3+1}
	\spab{m_2}.{q}.{\xi}^4
	,\nonumber\\
	\tilde{A}_n^{(7),-+}(q;m_2,m_3) &=&
	\spa{m_2}.{1}^3
	\spa{m_2}.{m_3}
	\spab{m_3+1}.{q}.{\xi}
	\spab{n}.{q}.{\xi}^3,
	\\
	\tilde{A}_n^{(1),+-}(q;m_2,m_3) &=&
	\spa{m_2}.{m_3}^3
	\spab{m_2}.{q}.{\xi}
	\spa{n}.{1}
	\spab{m_3+1}.{q}.{n}^3
	,\nonumber\\
	\tilde{A}_n^{(3),+-}(q;m_2,m_3) &=&
	\spab{m_2}.{q}.{\xi}^3
	\spa{m_2}.{1}
	\spab{n}.{q}.{\xi}
	\spa{m_3+1}.{m_3}^3
	,\nonumber\\
	\tilde{A}_n^{(5),+-}(q;m_2,m_3) &=&
	\spa{1}.{n}
	\spa{m_3}.{m_3+1}^3
	\spab{m_2}.{q}.{\xi}^4
	,\nonumber\\
	\tilde{A}_n^{(7),+-}(q;m_2,m_3) &=&
	\spa{m_2}.{1}
	\spa{m_2}.{m_3}^3
	\spab{m_3+1}.{q}.{\xi}^3
	\spab{n}.{q}.{\xi},
\\
	\tilde{A}_n^{(1),++}(q;m_2,m_3) &=&
	\spa{m_2}.{m_3}^3
	\spab{m_2}.{q}.{\xi}
	\spa{n}.{1}
	\spab{n}.{q}.{\xi}^2
	\spab{m_3+1}.{q}.{\xi}
	,\nonumber\\
	\tilde{A}_n^{(3),++}(q;m_2,m_3) &=&
	\spa{m_2}.{1}
	\spab{m_2}.{q}.{\xi}^3
	\spab{n}.{q}.{\xi}
	\spa{n}.{m_3}^2
	\spa{m_3+1}.{m_3}
	,\nonumber\\
	\tilde{A}_n^{(5),++}(q;m_2,m_3) &=&
	\spa{1}.{n}
	\spa{n}.{m_3}^2
	\spa{m_3}.{m_3+1}
	\spab{m_2}.{q}.{\xi}^4
	,\nonumber\\
	\tilde{A}_n^{(7),++}(q;m_2,m_3) &=&
	\spa{m_2}.{1}
	\spa{m_2}.{m_3}^3
	\spab{m_3+1}.{q}.{\xi}
	\spab{n}.{q}.{\xi}^3.
\end{eqnarray}

Eqs.~(\ref{eq:4qnmhv1}) and (\ref{eq:4qnmhv2}) are sufficient to describe all amplitudes involving $\phi$, two pairs of quarks and
a single negative helicity gluon.  Amplitudes involving $\phi^\dagger$ are obtained by parity. 
Note that all NMHV amplitudes lie in the overlap of the MHV and \MHVb\ towers.

The only NMHV amplitudes previously available were those involving
for four quarks and a single gluon~\cite{DFM}.

\subsubsection{$H\to q^{\lambda_1}g^-\bar{Q}^{-\lambda_2}Q^{\lambda_2}\bar{q}^{-\lambda_1}$}

In this case quarks of opposite flavour are colour connected corresponding to the
leading colour $A_{n}^{\lambda_1\lambda_2}$ NMHV with $m_2=2$ and $m_3=3$.  
To recover the amplitude for Higgs we add the \MHVb\ amplitude  with the same colour and helicity configuration,
\begin{eqnarray}
	A_5(H,1^{\lambda_1}_q,2^-,3^{-\lambda_2}_{\bar{Q}},4^{\lambda_2}_{Q},5^{-\lambda_1}_{\bar{q}}) &=& 
	A_5(\phi,1^{\lambda_1}_q,2^-,3^{-\lambda_2}_{\bar{Q}},4^{\lambda_2}_{Q},5^{-\lambda_1}_{\bar{q}})\nonumber\\
	&+& A_5(\phi^\dagger,1^{\lambda_1}_q,2^-,3^{-\lambda_2}_{\bar{Q}},4^{\lambda_2}_{Q},5^{-\lambda_1}_{\bar{q}}).
\end{eqnarray}
By substituting in specific phase space points with various choices of the gauge vector $\xi$,
we find numerically that the amplitude is gauge invariant.   We also find agreement with 
the results of an independent calculation of the twelve Feynman diagrams.

\subsubsection{$H\to q^{\lambda_1}g^-\bar{q}^{-\lambda_1}Q^{\lambda_2}\bar{Q}^{-\lambda_2}$}

In this case, each quark is colour connected to the antiquark of the same flavour.
We therefore take the subleading colour $\tilde{A}_{n}^{\lambda_1\lambda_2}$ NMHV with $m_2=2$ and $m_3=3$ and add
the \MHVb~ with the same configuration
\begin{eqnarray}
	A_5(H,1^{\lambda_1}_q,2^-,3^{-\lambda_1}_{\bar{q}},4^{\lambda_2}_{Q},5^{-\lambda_2}_{\bar{Q}}) &=& 
	\tilde{A}_5(\phi,1^{\lambda_1}_q,2^-,3^{-\lambda_1}_{\bar{q}},4^{\lambda_2}_{Q},5^{-\lambda_2}_{\bar{Q}})\nonumber\\
	&+& \tilde{A}_5(\phi^\dagger,1^{\lambda_1}_q,2^-,3^{-\lambda_1}_{\bar{q}},4^{\lambda_2}_{Q},5^{-\lambda_2}_{\bar{Q}}).
\end{eqnarray}
Once again, we find that the amplitude is gauge invariant and reproduces the numerical result found
using an independent Feynman diagram calculation.

\begin{figure}[t]
	\centering
	\psfrag{i+1}{\scriptsize$i+1$}
	\psfrag{j+1}{\scriptsize$j+1$}
	\psfrag{i}{\scriptsize$i$}
	\psfrag{j}{\scriptsize$j$}
	\psfrag{P}{\scriptsize$P$}
	\psfrag{a}{\scriptsize$a$}
	\psfrag{b}{\scriptsize$b$}
	\psfrag{c}{\scriptsize$c$}
	\includegraphics[width=0.60\textwidth]{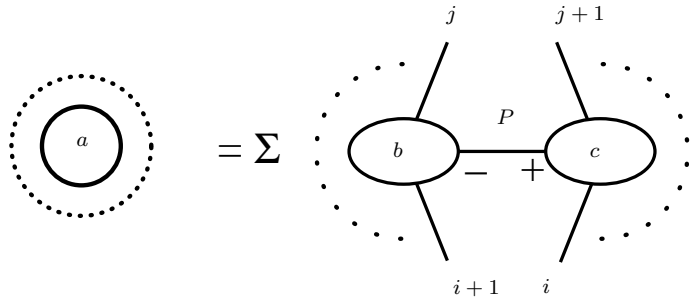}
	\caption{The recursion relation for amplitudes involving $a$ negative helicity particles. The dots indicate the
	emission of particles
	of any helicity.  $a$, $b$ and $c$ count the number of negative helicity external particles connected to each vertex, such that $a = b+c$.  The
	summation runs over all possible distributions of external particles. $P$ denotes the off-shell momentum linking the two
	vertices. }
	\label{fig:recur}
\end{figure}

\section{Recursive formulation of non-MHV amplitudes \label{sec:recurs}}

The NMHV amplitudes of the previous sections were obtained by connecting two MHV vertices by
a scalar propagator in all possible ways.   Typically there are of order 10 such diagrams.   
NNMHV amplitudes can be constructed either by connecting three MHV vertices, 
or by connecting an MHV vertex to an on-shell NMHV amplitude.   The first approach involves around 50 scalar graphs, 
while the second method recursively makes use of previously computed results.      
Recursion relations were first used in the context of QCD amplitudes by Berends and Giele~\cite{BerendsGiele}
and subsequently by Kosower~\cite{LightConeRecursive}.   More recently, they have been employed to 
obtain helicity amplitudes for gluon scattering using MHV rules~\cite{BBK}.

Following \cite{BBK}, an $n$-gluon amplitude with $a$ negative helicities, $A^g_{n;a}$, can be written in terms
of amplitudes involving fewer external particles with $b$ and $c$ negative helicities as,
\begin{align}
\label{eq:gluonrecur}
	A^{g}_{n;a}&(1,\ldots,n) =  \frac{1}{(a-2)}
	\sum_{i=1}^{n} \sum_{j=i+2}^{i-2}
	\nonumber \\
	&
	A^{g}_{j-i+1;b+1}(i+1,\ldots,j,-q_{i+1,j}^-) \,
	\frac{1}{s_{i+1,j}}\, A^{g}_{i-j+1;c}(j+1,\ldots,i,-q_{j+1,i}^+),
\end{align}
where $a = b+c$.
This relation is schematically shown in figure~\ref{fig:recur}.
The helicities of individual particles have been suppressed, however, it is understood that out of the $n$ particles, 
$a$ have negative helicity.   Furthermore, $b$ ($c$) particles in the range $i+1,\ldots,j$ $(j+1,\ldots,i)$ have negative
helicities.  All momenta are outgoing, and the offshell line linking the two
amplitudes carries momentum $q_{i+1,j} = -q_{j+1,i}$,
such that $s_{i+1,j}=q_{i+1,j}^2$,
and we choose to connect it to the 
left-(right-) hand amplitude with negative (positive) helicity.  
Note that the range $[i+1,j]$ must contain at least 1 negative helicity and the range $[j+1,i-1]$ must
contain at least 2 negative helicities.
Vertices generated which do not satisfy these properties will be
zero. It is also understood that all sums and ranges are defined modulo
$n$.   The factor ${1\over (a-2)}$ makes sure that there is no overcounting of diagrams. 

For our present purposes, we wish to use eq.~(\ref{eq:gluonrecur}) algebraically to reduce the amplitude to a combination 
of MHV vertices $A^g_{n;2}$ with some number of off-shell legs.  The off-shell continuation is then performed  
in the usual manner~\cite{CSW1},
$
\spa{i}.{P} \to \langle i | \slash \!\!\! P | \xi\rangle.
$
In this case, it is convenient to treat all external particles as on-shell and treat the off-shell legs analytically.
For purely numerical evaluation, it may be simpler to deal with vertices with
all legs off-shell at the beginning and take them on-shell afterwards~\cite{BBK}.  The off-shell continuation of Kosower
~\cite{KosowerNMHV} is particularly well suited to the numerical approach.  
 
We must now extend the recursive formula of ref.~\cite{BBK} to include both fermions and scalars. 
The relevant building blocks for amplitudes with up to one quark pair and/or 
one $\phi$ are thus the four MHV vertices, $A^g_{n;2}$, $A^q_{n;2}$,  $A^{\phi g}_{n;2}$ and
$A^{\phi q}_{n;2}$.  We do not indicate whether the quark or antiquark has negative helicity,
and so the quark MHV vertices are represented by a
single $A$.

We first write down the recursion relation for amplitudes involving $\phi$ and gluons,
\begin{align}
\label{eq:phirecur}
	A^{\phi g}_{n;a}&(1,\ldots,n) = \frac{1}{(a-2)}
	\bigg \{
	\nonumber\\
	&
	\sum_{i=1}^{n} \sum_{j=i+2}^{i-1}
	A^{g}_{j-i+1;b+1}(i+1,\ldots,j,-q_{i+1,j}^-) \,
	\frac{1}{s_{i+1,j}}\,
	A^{\phi g}_{i-j+1;c}(j+1,\ldots,i,q_{i+1,j}^+) 
	\nonumber\\
	+&
	\sum_{i=1}^{n} \sum_{j=i+1}^{i-2}
	A^{\phi g}_{j-i+1;b+1}(i+1,\ldots,j,q_{j+1,i}^-)\, 
	\frac{1}{s_{j+1,i}}\,
	A^{g}_{i-j+1;c}(j+1,\ldots,i,-q_{j+1,i}^+),
	\bigg \}.
\end{align}
Note that for amplitudes involving $\phi$, the outgoing 
gluon momenta no longer sum to
zero.  We therefore choose to use the momentum constructed solely from gluon
momenta, $q_{i+1,j}$ ($q_{j+1,i}$)  in the $\phi-$gluon vertex
appearing on first (second) terms on the rhs of eq.~(\ref{eq:phirecur}).
This expression is sufficient to rederive the NMHV and Next-to-NMHV (NNMHV) amplitudes
given explicitly in Ref.~\cite{DGK}.  We have checked that it correctly gives
NMHV and NNMHV amplitudes with up to six gluons.\newpage

The recursion relation involving only quarks and gluons is given by,
\begin{align}
\label{eq:qrecur}
	A^{q}_{n;a}&(1_q^\lambda,\ldots,n^{-\lambda}_{\bar{q}}) 
	= \frac{1}{(a-2)}
	\Bigg ( \nonumber \\
	&\sum_{i=1}^{n-3}\sum_{j=i+2}^{n-1} \bigg \{
	\nonumber\\
	&
	A^{g}_{j-i+1;b+1}(i+1,\ldots,j,-q_{i+1,j}^-)\,
	\frac{1}{s_{i+1,j}}\,
	A^{q}_{i-j+n+1;c}(1_q^\lambda,2,\ldots,i,-q_{j+1,i}^+,j+1,\ldots,n_{\bar{q}}^{-\lambda})
	\nonumber\\
	+&
	A^{g}_{j-i+1;b}(i+1,\ldots,j,-q_{i+1,j}^+)\,
	\frac{1}{s_{i+1,j}}\,
	A^{q}_{i-j+n+1;c+1}(1_q^\lambda,2,\ldots,i,-q_{j+1,i}^-,j+1,\ldots,n_{\bar{q}}^{-\lambda})
	\bigg \} \nonumber\\
	+& \sum_{i=2}^{n-2}
	A^{q}_{i+1;b^\prime+1}(1_q^\lambda,2,\ldots,i,-q_{1,i}^{-\lambda})
	\frac{1}{s_{1,i}}
	A^{q}_{n-i+1;c^\prime+1}(q_{1,i}^{\lambda},i+1,\ldots,n_{\bar{q}}^{-\lambda})
	\Bigg ),
\end{align}
where $b^\prime$ ($c^\prime$) 
is the number of negative helicities in the range $2,\ldots,i$ ($i+1,\ldots,
n-1$) and $a = b^\prime+c^\prime-1$.
Eq.~(\ref{eq:qrecur}) is sufficient to describe amplitudes involving any number
of gluons and a single quark pair.

Finally, the recursion relation for quarks, gluons and a $\phi$ is,
\begin{align}
\label{eq:phiqrecur}	%
	A^{\phi q}_{n;a}&(1_q^\lambda,\ldots,n^{-\lambda}_{\bar{q}}) = \frac{1}{(a-2)}
	\Bigg ( 
	\nonumber\\
	&\sum_{i=1}^{n-2}\sum_{j=i+1}^{n-1} 
	\bigg \{
	\nonumber\\
	&
	A^{\phi g}_{j-i+1;b+1}(i+1,\ldots,j,q_{j+1,i}^-)\,
	\frac{1}{s_{j+1,i}}\,
	A^{q}_{i-j+n+1;c}(1_q^\lambda,2,\ldots,i,-q_{j+1,i}^+,j+1,\ldots,n_{\bar{q}}^{-\lambda})
	\nonumber\\
	+&
	A^{\phi g}_{j-i+1;b}(i+1,\ldots,j,q_{j+1,i}^+)\,
	\frac{1}{s_{j+1,i}}\,
	A^{q}_{i-j+n+1;c+1}(1_q^\lambda,2,\ldots,i,-q_{j+1,i}^-,j+1,\ldots,n_{\bar{q}}^{-\lambda})
	\bigg \}
	\nonumber\\
	+&\sum_{i=1}^{n-3}\sum_{j=i+2}^{n-1} 
	\bigg \{
	\nonumber\\
	&
	A^{g}_{j-i+1;b+1}(i+1,\ldots,j,-q_{i+1,j}^-)\,
	\frac{1}{s_{i+1,j}}\,
	A^{\phi q}_{i-j+n+1;c}(1_q^\lambda,2,\ldots,i,q_{i+1,j}^+,j+1,\ldots,n_{\bar{q}}^{-\lambda})
	\nonumber\\
	+&
	A^{g}_{j-i+1;b}(i+1,\ldots,j,-q_{i+1,j}^+)\,
	\frac{1}{s_{i+1,j}}\,
	A^{\phi q}_{i-j+n+1;c+1}(1_q^\lambda,2,\ldots,i,q_{i+1,j}^-,j+1,\ldots,n_{\bar{q}}^{-\lambda})
	\bigg \} \nonumber\\
	+& \sum_{i=2}^{n-2}
	\bigg \{
	\nonumber\\
	&
	A^{\phi q}_{i+1;b^\prime+1}(1_q^\lambda,2,\ldots,i,q_{i+1,n}^{-\lambda})\,
	\frac{1}{s_{i+1,n}}\,
	A^{q}_{n-i+1;c^\prime+1}(-q_{i+1,n}^{\lambda},i+1,\ldots,n_{\bar{q}}^{-\lambda})
	\nonumber\\
	+&
	A^{q}_{i+1;b^\prime+1}(1_q^\lambda,2,\ldots,i,-q_{1,i}^{-\lambda})\,
	\frac{1}{s_{1,i}}\,
	A^{\phi q}_{n-i+1;c^\prime+1}(q_{1,i}^{\lambda},i+1,\ldots,n_{\bar{q}}^{-\lambda})
	\bigg \}
	\Bigg ).  
\end{align}
We have checked that eq.~(\ref{eq:phiqrecur}) correctly reproduces the NMHV amplitudes given in section 3
for up to 6 gluons.
Note that in order for the recursion relation to be effective, and unlike the case for the 
explicit formulae for NMHV amplitudes in eq.~(\ref{eq:NMHV}) that is valid for all $n$, 
the number of particles must be specified.
The recursion relation and the explicit all $n$ results are therefore complementary.

Equations (\ref{eq:gluonrecur})--(\ref{eq:phiqrecur})
provide a way to generate expressions for all non-MHV amplitudes with fermions, gluons
and a single massive scalar, $\phi$.  As usual, amplitudes involving $\phi^\dagger$
are obtained by parity. 

\subsection{$H \to q^-g^-g^-g^-\bar{q}^+$}
The only NNMHV amplitude previously available in the literature is for
$H \to q^-g^-g^-g^-\bar{q}^+$~\cite{DFM}.  This corresponds to the point with $n_++n_-=5$ and
$n_+-n_-=-3$ in figure~\ref{fig:treemap}.  There is no \MHVb\ 
contribution with this helicity configuration.
The full amplitude is thus,
\begin{eqnarray}
	A_n(H,q^-_1,g^-_2,g^-_3,g^-_4,\bar{q}^+_5) &=&
	A_n(\phi,q^-_1,g^-_2,g^-_3,g^-_4,\bar{q}^+_5).
\end{eqnarray}
We have checked numerically that the amplitude obtained using the
recursion relation (\ref{eq:phiqrecur}) is gauge invariant 
and that it agrees with the results obtained by computing the 74 Feynman diagrams directly.

\section{Conclusions}
\label{ConclusionSection}

In this paper we have presented a
generalised set of MHV rules for constructing perturbative amplitudes involving a massive
Higgs boson plus an arbitrary number of partons -- quarks and gluons.
These MHV rules incorporate interactions with quarks into the recent construction of
\cite{DGK} for amplitudes for a Higgs and any number of gluons.

We use an effective dimension 5 interaction to approximate the top quark mediated one-loop coupling
of the Higgs field to the gluons. This effective vertex is added to the standard QCD
Lagrangian. The resulting effective theory is used at tree level to generate the MHV rules
for computing perturbative amplitudes. As in Ref. \cite{DGK} the MHV rules are generated by
splitting the effective interaction $HG_{\mu\nu}G^{\mu\nu}$ into
selfdual and anti-selfdual pieces that generate the two towers of MHV and \MHVb\ diagrams
that are crucial for the construction to work. 
Adding quarks to the formalism of \cite{DGK}
generates additional building blocks in the MHV perturbation theory --
the vertices coupling one and two
quark-antiquark pairs to the Higgs and gluons. These
additional MHV (\MHVb\ ) vertices are uniquely determined from the 
purely gluonic $\phi$-MHV ($\phi^\dagger$-\MHVb\
vertices of \cite{DGK} via a (fake) supersymmetry.

Our MHV rules lead
to compact formulae for the Higgs plus multi-parton amplitudes
induced at leading order in QCD in the large $m_t$ limit.
We presented explicit formulae for the Next-to-MHV $\phi$-plus-$n$-parton amplitudes
and checked that for $n\le 5$ these results agree with the expressions derived from
Feynman diagrams.

We have also shown how to apply the MHV rules recursively to get $n$-particle amplitudes with
arbitrary numbers of negative and positive helicities, for processes involving quarks, gluons and
massive scalars. This is achieved by generalising the recursion relations for MHV rules
from the purely gluonic amplitudes of Bena, Bern and Kosower \cite{BBK}, to the case with
Higgs, fermions and gluons. We used this recursive formulation of MHV rules to
numerically check Next-to-Next-to-MHV amplitudes with the Higgs and 5 partons.
Of course the recursion relations are valid for any number of partons.

\acknowledgments 

We thank Vittorio Del Duca, Alberto Frizzo and Fabio Maltoni 
for providing us with
a numerical program computing the amplitudes of ref.~\cite{DFM}.  We thank Lance Dixon for helpful discussions
at the outset of this work.
EWNG and VVK acknowledge the support of PPARC through 
Senior Fellowships and SDB acknowledges the award of a PPARC studentship.

\newpage
\appendix


\section{Conventions}

We work in Minkowski space with the metric
$\eta^{\mu\nu}$  and use the sigma matrices from
Wess and Bagger~\cite{WessBagger},
$\sigma^\mu_{\alpha \dot\alpha}=(-1,\tau^1,\tau^2,\tau^3)$, and
$(\bar\sigma^{\mu})^{\dot\alpha \alpha}=(-1,-\tau^1,-\tau^2,-\tau^3)$, where
$\tau^{1,2,3}$ are the Pauli matrices.

In the spinor helicity formalism
\cite{SpinorHelicity1,SpinorHelicity2,SpinorHelicity3,SpinorHelicity4,SpinorHelicity5,SpinorHelicity6} an on-shell momentum
of a massless particle, $k_\mu k^\mu=0$, is represented as
\be
k_{\alpha \dot\alpha} \equiv \ k_\mu \sigma^\mu_{\alpha \dot\alpha}
=\ \lambda_\alpha\tilde\lambda_{\dot\alpha} \ ,
\ee
where $\lambda_\alpha$ and $\tilde\lambda_{\dot\alpha}$
are two commuting spinors of positive and negative chirality.
Spinor inner products are defined
by\footnote{Our conventions for spinor helicities follow
refs.~\cite{Witten1,CSW1}, except that $[ij] = - [ij]_{CSW}$
as in ref.~\cite{LDTASI}.}
\be
\langle \lambda,\lambda'\rangle 
= \ \epsilon_{\alpha\beta}\lambda^\alpha\lambda'{}^\beta
 \,, \qquad
[\tilde\lambda,\tilde\lambda'] 
=\ -\epsilon_{\dot\alpha \dot\beta}
\tilde\lambda^{\dot\alpha}\tilde\lambda'{}^{\dot\beta} \,,
\label{Atwo}
\ee
and a scalar product of two null vectors,
$k_{\alpha \dot\alpha}=\lambda_\alpha \tilde\lambda_{\dot\alpha}$ and
$p_{\alpha \dot\alpha}=\lambda'_\alpha\tilde\lambda'_{\dot\alpha}$, becomes
\be \label{scprod}
k_\mu p^\mu =\ - {1\over 2}
\langle\lambda,\lambda'\rangle[\tilde\lambda,\tilde\lambda'] \,.
\ee
We use the shorthand $\spa{i}.{j}$ and $\spb{i}.{j}$ for the inner products of
the spinors corresponding to momenta $p_i$ and $p_j$,
\be
\spa{i}.{j} = \langle \lambda_i, \lambda_j \rangle \,,
\qquad
\spb{i}.{j} = [ \tilde\lambda_i, \tilde\lambda_j ].
\ee

For gluon polarization vectors we use
\be
\pol_\mu^\pm(k,\xi) = \pm { \langle \xi^\mp | \gamma_\mu | k^\mp \rangle
                 \over \sqrt{2} \langle \xi^\mp | k^\pm \rangle } \,,
\label{HelPol}
\ee
where $k$ is the gluon momentum and $\xi$ is the reference momentum, an
arbitrary null vector which can be represented as the product of two
reference spinors, 
$\xi_{\alpha\dot\alpha}=\xi_{\alpha}\tilde\xi_{\dot\alpha}$. 
We choose the reference momenta for all gluons to be the same, unless 
otherwise specified.  In terms of helicity spinors, 
$\pol_{\alpha \dot\alpha} = \pol^\mu (\gamma_\mu)_{\alpha \dot\alpha}$,
\eqn{HelPol} takes the form \cite{Witten1},
\bea
\pol_{\alpha \dot\alpha}^+ &=&
\sqrt{2} \frac{\xi_\alpha \tilde\lambda_{\dot\alpha}}
{\vev{\xi~\lambda}} \,,
\label{Pluspolvhel} \\ 
\pol_{\alpha \dot\alpha}^{-} &=& 
\sqrt{2} \frac{\lambda_\alpha \tilde{\xi}_{\dot\alpha}}
{[\tilde\lambda ~ \tilde{\xi}]} \,.
\label{Minuspolvhel}
\eea
To simplify the notation, we will drop the tilde-sign over the dotted 
reference spinor, so that 
$\xi_{\alpha \dot\alpha} = \xi_{\alpha}\xi_{\dot\alpha}$.

\section{Colour decomposition} \label{Cdecomp}

In general, an $n$-point amplitude ${\cal A}_n $
can be represented as a sum of products of colour factors ${\cal T}_n$
and purely kinematic partial
amplitudes $A_n$,
\be
{\cal A}_n (\{p_i,\lambda_i,c_i\}) \,=\, \sum_{\sigma} \,
{\cal T}_n (\{c_{\sigma(i)}\}) \, A_n (\{p_{\sigma(i)},\lambda_{\sigma(i)}\}) \, .
\label{one}
\ee
Here $\{c_i\}$ are colour labels of external legs $i=1 \ldots n$, and
the kinematic variables $\{p_i,\lambda_i\}$ are on-shell external momenta and helicities:
all $p_i^2=0,$ and $\lambda_i=\pm 1$ for gluons, $\lambda_i=\pm {1\over 2}$ for fermions, and
$\lambda_i=0$ for scalars.
The sum in \eqref{one} is over appropriate simultaneous permutations $\sigma$ of
colour labels $\{c_{\sigma(i)}\}$ and kinematic variables
$\{k_{\sigma(i)},\lambda_{\sigma(i)}\}$. To simplify expressions, we will often denote
$p_i,\lambda_i$ as $p_i^{\lambda_i}$.

The tree-level Higgs-plus-gluons amplitudes can be decomposed into
color-ordered partial amplitudes~\cite{DawsonKauffman,DFM} as
\begin{equation}
{\cal A}_n(H,\{p_i,\lambda_i,a_i\}) =
i C g^{n-2}
\sum_{\sigma \in S_n/Z_n} \Tr(T^{a_{\sigma(1)}}\cdots T^{a_{\sigma(n)}})\,
A_n(H,\sigma(1^{\lambda_1},\ldots,n^{\lambda_n}))\,.
\label{TreeColorDecomposition}
\end{equation}
Here $S_n/Z_n$ is the group of non-cyclic permutations on $n$
symbols, and $j^{\lambda_j}$ labels the momentum $p_j$ and helicity
$\lambda_j$ of the $j^{\rm th}$ gluon, which carries the adjoint
representation index $a_i$.  The $T^{a_i}$ are fundamental
representation SU$(N_c)$ color matrices, normalized so that
$\Tr(T^a T^b) = \delta^{ab}$.  The strong coupling constant is
$\alpha_s=g^2/(4\pi)$.

Color-ordering means that, in a computation based on Feynman diagrams,
the partial amplitude
$A_n(H,1^{\lambda_1},2^{\lambda_2},\ldots,n^{\lambda_n})$
would receive contributions only from planar tree diagrams with a
specific cyclic ordering of the external gluons: $1,2,\ldots,n$.
Because the Higgs boson is uncolored, there is no color restriction on how
it is emitted.  The partial amplitude $A_n$ is invariant under
cyclic permutations of its gluonic arguments.

Two comments are in order: first, is that \eqref{TreeColorDecomposition} is valid
for all tree-level amplitudes involving a colourless Higgs plus any fields in
the adjoint representation (e.g. gluons, gluinos but no fundamental quarks).
Second comment is that because the Higgs carries no colour, it does not enter into
the colour factor and the decomposition
 \eqref{TreeColorDecomposition} can also be applied to pure
(supersymmetric and non-supersymmetric) Yang-Mills without
the Higgs. To do this, we simply remove the Higgs field $H$ from both sides of
\eqref{TreeColorDecomposition} and set $C=1$.

Fields in the fundamental representation, in particular quarks and anti-quarks,
are included in the colour-ordered amplitudes as follows.
For a fixed colour ordering $\sigma$, the amplitude
with $m$ quark-antiquark pairs and $l$ gluons (and gluinos)
is given by
\be
{\cal T}_{l+2m} (\{c_{\sigma(i)}\}) \ A_{l+2m} (H,\{p_{\sigma(i)},\lambda_{\sigma(i)}\}) \, .
\label{three}
\ee
(The full amplitude \eqref{one} is the sum over all appropriate permutations
$\sigma$ in \eqref{three}.)
For tree amplitudes
the exact colour factor in
\eqref{three} is \cite{SpinorHelicity6}
\be
T_{l+2m} \, = \,
{(-1)^p\over N^p} ({\rm T}^{a_1} \ldots {\rm T}^{a_{l_1}})_{i_1 \alpha_1}
({\rm T}^{a_{l_1+1}} \ldots {\rm T}^{a_{l_2}})_{i_2 \alpha_2} \ldots
({\rm T}^{a_{l_{m-1}+1}} \ldots {\rm T}^{a_{l}})_{i_m \alpha_m} \, .
\label{cfqq}
\ee
Here $l_1, \ldots , l_{m}$ correspond to an arbitrary partition of an arbitrary
permutation of the $l$ gluon (and gluino) indices; $i_1, \ldots i_m$ are
colour indices of quarks, and $\alpha_1, \ldots \alpha_m$ -- of the antiquarks.
In perturbation theory each external quark is connected by a fermion line to an external
antiquark (all particles are counted as incoming). When quark $i_k$ is connected by a fermion
line to antiquark $\alpha_k$, we set $\alpha_k=\bar{i_k}$. Thus, the set of
$\alpha_1, \ldots \alpha_m$ is a permutation of the set
$\bar{i_1}, \ldots \bar{i_m}$. Finally,
the power $p$ is equal to the number of times $\alpha_k=\bar{i_k}$ minus 1.
When there is only one quark-antiquark pair, $m=1$ and $p=0$. For a general $m$,
the power $p$ in \eqref{cfqq} varies from $0$ to $m-1$.

The kinematic amplitudes $A_{l+2m}$ in \eqref{three} have the colour information stripped off
and hence do not distinguish between fundamental quarks and adjoint gluinos.
Hence, if we know kinematic amplitudes involving gluinos in a supersymmetric theory,
we automatically know kinematic amplitudes with quarks,
\be
A_{l+2m}(H,q^{+},\ldots,\bar{q}^{-},\ldots, g^+,\ldots, g^-)\, = \,
A_{l+2m}(H,\Lambda^+,\ldots,\Lambda^-,\ldots, g^+,\ldots, g^-)
\label{four}
\ee
Here $q^{\pm}$, $\bar{q}^{\pm}$, $g^{\pm}$, $\Lambda^{\pm}$ denote
quarks, antiquarks, gluons and gluinos of $\pm$ helicity.

Since the Higgs carries no colour,
it does not enter into \eqref{cfqq}, and
\eqref{four} is valid with or without the Higgs field.
We conclude from \eqref{four} that knowing kinematic amplitudes in a supersymmetric
theory with gluinos allows us to deduce immediately non-supersymmetric amplitudes with quarks
and anti-quarks.

The only difference between (anti)~quarks and gluinos is that (anti)~quarks, being
(anti)~fundamental fields, can enter the colour factors \eqref{cfqq}
and hence the kinematic amplitudes, only in specific places, dictated
by the ${i_k \alpha_k}$ indices in \eqref{cfqq}. Gluinos, on the other hand,
transform in the adjoint, and can enter colour-ordered amplitudes in arbitrary places.

\newpage
\section{Vanishing of $A_n(\phi,q^\lambda_1,g^+_2,\ldots,g^+_{n-1},\bar{q}^{-\lambda}_n)$ \label{app:vanish}}

In this appendix we demonstrate that the $A_n(\phi,q^\lambda_1,g^+_2,\ldots,g^+_{n-1},\bar{q}^{-\lambda}_n)$
amplitude vanishes using the traditional Feynman approach and Berends-Giele currents.
We find that the proof proceeds in almost exactly the same way as the case for gluon amplitudes (see
Appendix B2 of reference \cite{DGK}):
\begin{equation}
	A_n(\phi,g_1^\pm,g_2^+,\ldots,g_n^+) = 0.
\end{equation}

\begin{figure}[h!]
	\psfrag{n}{$n$}
	\psfrag{n-1}{$n-1$}
	\psfrag{i}{$m-1$}
	\psfrag{1}{$1$}
	\psfrag{2}{$2$}
	\psfrag{i+1}{$m$}
	\psfrag{phi}{$\phi$}
	\psfrag{S}{$S^\mu _+$}
	\psfrag{J}{$J^\mu _+$}
	\begin{center}
		\includegraphics[width=7cm]{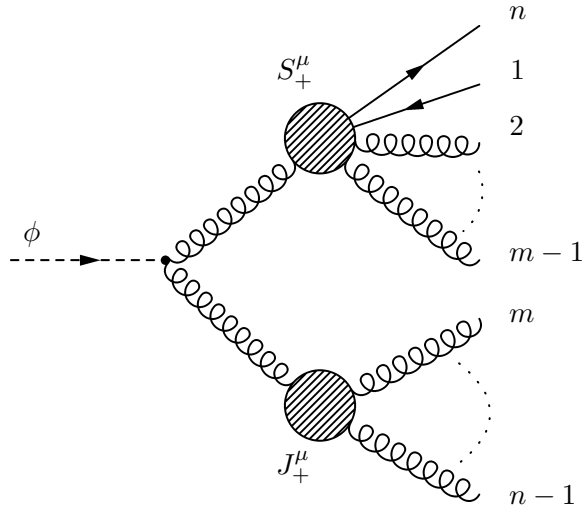}
	\end{center}
	\caption{The contribution to $A_n(\phi,q^-_1,g^+_2,\ldots,g^+_{n-1},\bar{q}^+_n)$ coming from
	a quark ``all plus" current and a gluon ``all plus" current joined by a $\phi gg$ vertex.}
	\label{fig:current}
\end{figure}

In our case we need to consider two currents. The first consists off an off-shell gluon attached to any
number of positive helicity gluons, denoted $J_+^\mu$, and the second consists of an off-shell gluon
attached to a quark pair and any number of positive helicity gluons, denoted $S_+^\mu$. The relevant diagram 
is shown in figure~\ref{fig:current}. It is important to notice that the scalar $\phi$ can only couple to gluons in our
effective model and hence there is no $\phi \to q\bar{q}$ vertex. For simplicity we will set the quark helicity to be negative,
$\lambda=-1$. The case where $\lambda=+1$ follows by an identical calculation. 

The 3-point $\phi-gg$ vertex is given by:
\begin{equation}
V^\phi_{\mu_1\mu_2}(p_1,p_2) =
\eta_{\mu_1\mu_2} p_1 \cdot p_2 - p_{1\,\mu_2} p_{2\,\mu_1}
 + i \pol_{\mu_1\mu_2\nu_1\nu_2} p_1^{\nu_1} p_2^{\nu_2} \,,
\label{eq:Hggvert}
\end{equation}
where $\mu_i$ are Lorentz indices and $p_1$ and $p_2$ are the outgoing momenta. 

We can compute the case of $A_3(\phi,q^-_1,g_2^\pm,\bar{q}^+_3)$ very simply by contracting the vertex
(\ref{eq:Hggvert}) with a polarisation vector $\e_\pm^\mu(p_2)$ and a quark-antiquark current
$\epsilon_q^\nu(k) = \la {1-} | \gamma^\nu | {3-} \ra/s_{13}$ where $k=p_1+p_3$. Just as in \cite{DGK} we compute the
ratio of the 3rd to 2nd term in equation (\ref{eq:Hggvert}) (the 1st term gives zero):
\begin{eqnarray}
\label{eq:r3def}
	R^+_3 =
	\frac{i\epsilon_{\mu_1\mu_2\nu_1\nu_2}\e^{\mu_1}(p_2)\e^{\mu_2}_q(k)p_2^{\nu_1}k^{\nu_2}}
	{-\e_+(p_2)\cdot p_1 \e_q(k)\cdot p_2}.
\end{eqnarray}
Using the identities:
\begin{eqnarray}
	i\e_{\mu_1\mu_2\mu_3\mu_4} &=&
	\frac{1}{4}\rm{tr}(\gamma_5\gamma^{\mu_1}\gamma^{\mu_2}\gamma^{\mu_3}\gamma^{\mu_4})
	\label{eq:ids1},\\
	\tfrac{1}{2}(1\pm\gamma_5)\gamma_\mu\langle a\pm | \gamma^\mu | b\pm \rangle &=&
	2|b\pm\rangle\langle a\pm | ,\label{eq:ids2}
\end{eqnarray}
it is easy to show $R^+_3 = -1$ and therefore $A_3(\phi,q^-_1,g_2^+,\bar{q}^+_3) = 0$.
Similarly, we can show that $R^-_3=+1$ and that the sum of the two terms does indeed match the
proposed form of equation~(\ref{eq:2qMHV2}).

In order to extend this method to prove that the $n$-particle amplitude vanishes we make use of the two
currents mentioned before:
\begin{eqnarray}
	J^\mu _+(1^+,\ldots,n^+) &=& \frac{\langle \xi^- | \gamma^\mu \Psl_{1,n} | \xi^+ \rangle}
	{\sqrt{2}\spa{\xi}.{1}\spa{1}.{2}\ldots\spa{n}.{\xi}} ,\\
	S^\mu _+(1_q^-,2^+,\ldots,n-1^+,n_{\bar{q}}^+) &=& \frac{\langle \xi^- | \gamma^\mu \Psl_{1,n-1} | \xi^+ \rangle}
	{\spa{\xi}.{2}\spa{2}.{3}\ldots\spa{n-1}.{\xi}}.
\end{eqnarray}
We immediately notice that the currents have a very similar form. Indeed since the denominators play
no role in making sure the amplitude vanishes, it is obvious that the proof will proceed in the same
way as the gluon case. So all that remains is to note:
\begin{eqnarray}
\label{JSvanish} 
	J_+ \cdot S_+ &=& 0 ,\\
\label{EpsJJSvanish}
	\e_{\mu\nu\rho\sigma}J_+^\mu J_+^\nu S_+^\sigma &=& 0 ,\\
\label{EpsJJJvanish}
	\e_{\mu\nu\rho\sigma}J_+^\mu J_+^\nu J_+^\sigma &=& 0.
\end{eqnarray}
These relations in turn suffice to show that the
Feynman vertices coupling $\phi$ to 3 or 4 gluons, $\phi ggg$ and $\phi gggg$,
do not contribute to $A_n(\phi,1^\pm,2^+,3^+,\ldots,n^+)$.
Terms in these vertices without a Levi-Civita tensor always attach
a Minkowski metric $\eta_{\mu_1\mu_2}$ to two currents; their
contribution vanishes according to \eqn{JSvanish}.
(The same is true of the first term in the $\phi gg$
vertex~(\ref{eq:Hggvert}).)
Terms containing the Levi-Civita tensor $\pol_{\mu_1\mu_2\mu_3\mu_4}$
attach it directly to at least three currents;
their contribution vanishes according to \eqns{EpsJJSvanish}{EpsJJJvanish}.
This leaves just the contributions of the second and third terms
in the $\phi gg$ vertex~(\ref{eq:Hggvert}).
They cancel against each other, just as in the case of
$A_3(\phi,q^-_1,g_2^+,\bar{q}^+_3)$ above.  Suppose that the quark current involving
gluons $2$ to $m-1$
is attached to one leg of the $\phi gg$ vertex, and that the current
involving gluons
$m$ to  $n-1$  is attached to the other leg.
Then the ratio analogous to \eqn{eq:r3def} is,
\begin{eqnarray}
\label{eq:rndef}
	R^+_n &=&
	\frac{i\epsilon_{\mu_1\mu_2\nu_1\nu_2}\langle \xi^- | \gamma^{\mu_1} \Psl_{1,m-1} | \xi^+ \rangle
	\langle \xi^- | \gamma^{\mu_2} \Psl_{m,n-1} | \xi^+ \rangle p_{1,m-1}^{\nu_1}p_{m,n-1}^{\nu_2}}
	{-\langle \xi^- | \Psl_{m,n-1} \Psl_{1,m-1} | \xi^+ \rangle \langle \xi^- | \Psl_{1,m-1} \Psl_{m,n-1} | \xi^+ \rangle}\\
	&=& -1.
\end{eqnarray}
This completes the recursive proof that,
\begin{equation}
	A_n(\phi,q^-_1,g^+_2,\ldots,g^+_{n-1},\bar{q}^{+}_n) = 0.
\end{equation}
A similar result holds when the quark has positive helicity.


\end{document}